\RequirePackage{xcolor}
\RequirePackage{ifpdf}
\documentclass[letterpaper]{JHEP3}
\usepackage{amsmath}
\usepackage{epsfig}
\usepackage{subfigure}
\usepackage{enumitem}
\pdfoutput=1

\newcommand{\roughly}[1]{\mathrel{\raise.3ex\hbox{$#1$\kern-0.85em
\lower1ex\hbox{$\sim$}}}}
\newcommand{\lsim}{\roughly<}
\newcommand{\gsim}{\roughly>}

\def\nn{\nonumber}

\newcommand{\be}{\begin{equation}}
\newcommand{\bee}{\begin{equation}}
\newcommand{\ee}{\end{equation}}
\newcommand{\beea}{\begin{eqnarray}}
\newcommand{\eea}{\end{eqnarray}}
\newcommand{\bea}{\begin{eqnarray}}

\def\nott#1{\setbox0=\hbox{$#1$}                
   \dimen0=\wd0                                 
   \setbox1=\hbox{/} \dimen1=\wd1               
   \ifdim\dimen0>\dimen1                        
      \rlap{\hbox to \dimen0{\hfil/\hfil}}      
      #1                                        
   \else                                        
      \rlap{\hbox to \dimen1{\hfil$#1$\hfil}}   
      /                                         
   \fi}                                         %

\def\uxsl{\hbox{/\kern-.4000em$u$}}
\def\uxslsm{\hbox{\smaller/\kern-.5600em$u$}}
\def\pxpsl{\hbox{/\kern-.5000em$p$}}
\def\epssl{\hbox{/\kern-.5600em$\epsilon$}}
\def\delsl{\hbox{/\kern-.7000em$\nabla$}}
\def\lxpsl{\hbox{/\kern-.5600em$l$}}
\def\kxpsl{\hbox{/\kern-.5600em$k$}}
\def\qxpsl{\hbox{/\kern-.3900em$q$}}

\def\pref#1{(\ref{#1})}
\def\exd{{\rm d}}
\def\ol#1{{\overline{#1}}}

\def\cA{{\cal A}}

\def\cD{{\cal D}}

\def\cH{{\cal H}}

\def\cL{{\cal L}}

\def\cO{{\cal O}}
\def\cP{{\cal P}}

\def\cR{{\cal R}}
\def\cS{{\cal S}}
\def\cT{{\cal T}}

\def\cV{{\cal V}}
\def\cW{{\cal W}}

\def\bfr{{\bf r}}

\def\mfa{{\mathfrak a}}
\def\mfb{{\mathfrak b}}
\def\mfc{{\mathfrak c}}

\def\mff{{\mathfrak f}}
\def\mfg{{\mathfrak g}}

\def\mft{{\mathfrak t}}

\def\mfD{{\mathfrak D}}

\def\mfK{{\mathfrak{K}}}

\def\ssA{{\scriptscriptstyle A}}
\def\ssB{{\scriptscriptstyle B}}
\def\ssC{{\scriptscriptstyle C}}

\def\ssG{{\scriptscriptstyle G}}
\def\ssH{{\scriptscriptstyle H}}
\def\ssI{{\scriptscriptstyle I}}
\def\ssJ{{\scriptscriptstyle J}}

\def\ssM{{\scriptscriptstyle M}}
\def\ssN{{\scriptscriptstyle N}}

\def\ssS{{\scriptscriptstyle S}}
\def\ssT{{\scriptscriptstyle T}}
\def\ssU{{\scriptscriptstyle U}}

\def\ssX{{\scriptscriptstyle X}}
\def\ssY{{\scriptscriptstyle Y}}

\def\KK{{\scriptscriptstyle KK}}

\def\GUT{{\scriptscriptstyle GUT}}

\title{RG-Induced Modulus Stabilization: Perturbative\\ de Sitter Vacua and  Improved D3-$\ol{\hbox{D3}}$ Inflation}

\author{C.P.~Burgess${}^{1,2,3}$ and F.~Quevedo${}^{4}$\\

{\it 
${}^1$ Department of Physics \& Astronomy, McMaster University, 
 Hamilton ON, Canada.\\
${}^2$ Perimeter Institute for Theoretical Physics, 
Waterloo ON, Canada.\\
${}^3$ CERN, Theoretical Physics Department, Gen\`eve 23, Switzerland.\\
${}^4$ DAMTP, Cambridge University, Wilberforce Road,  Cambridge, CB3 0WA, UK.
}
}

\preprint{CERN-TH-2022-009}
\date{\today}

\abstract{We propose a new mechanism that adapts to string theory a perturbative method for stabilizing moduli without leaving the domain of perturbative control, thereby evading the `Dine-Seiberg' problem. The only required nonperturbative information comes from the standard renormalization-group resummation of leading logarithms that allow us simultaneously to work to a fixed order in the perturbative parameter $\alpha$ and to all orders in $\alpha \ln\tau$ where $\tau$ is a large extra-dimensional modulus. The resulting potential is naturally minimized for moduli of order $\tau\sim e^{1/\alpha}$ and so can be exponentially large given  $\cO(10)$ input parameters. The mechanism relies on accidental low-energy scaling symmetries known to be generic  and so is robust against UV details. The resulting compactifications generically break supersymmetry and 4D de Sitter solutions are relatively easy to achieve without additional uplifting. Variations on the theme lead to inflationary scenarios for which the size of the stabilized moduli differ significantly before and after inflation and so provide a dynamical mechanism whereby inflationary scales are much larger than late-time physical ({\it e.g.}~supersymmetry breaking) scales, with this hierarchy contingent on past cosmic evolution with the inflaton playing a secondary late-time role as a relaxation field. 
We apply this formalism to warped D3-$\ol{\hbox{D3}}$ inflation using non-linearly realized supersymmetry to describe the antibrane tension and the Coulomb interaction, and show how doing so our perturbative modulus stabilization mechanism evades the $\eta$-problem that usually plagues this scenario. We speculate about the relevance of our formalism to tachyon condensation at later stages of brane-antibrane annihilation.  }

\begin{document}


\section{Introduction}

If string theory is right there may be a problem: a simple argument suggests string vacua are generically strongly coupled and without hierarchies, yet we find ourselves in a world populated by different scales where experiments often reveal weak interactions at play. How can these be reconciled? As it turns out, modulus stabilization is the key. Modulus stabilization is to string theory what logistics is to warfare: it is the difference between winning and losing. Our goal in this paper is to add a new stabilization mechanism to the string theory toolbox but we first start with a fuller statement of the issue and why our mechanism helps.

The issue turns about a central string-theory feature: scarcity of free parameters. Things that would be coupling constants in other theories arise as fields in string theory, making all expansions ultimately field expansions. For instance weak string coupling is an expansion in powers of the string dilaton $e^{\hat \phi} = 1/s$ and a 4D world only emerges from higher dimensions through an expansion in inverse powers of fields that express extra-dimensional size (such as the volume modulus $\cV:=\tau^{3/2}$ that measures its overall volume in string units). 

\subsubsection*{The `Dine-Seiberg' problem}

Crucially, fields like $s$ and $\cV$ are moduli, in that their values are often not fixed by the leading classical field equations. This means their vacuum values can be determined by energetic arguments purely within a four-dimensional low-energy effective theory through their appearance within a scalar potential that is to be minimized. Performing this stabilization is a prerequisite for making practical predictions because particle masses and couplings depend on the resulting stabilized values in an important way. However this potential itself typically arises as an expansion in these fields, as in
\be
V(s,\tau)=\sum_{nm} A_{nm} s^{-n} \tau^{-m} \,,
\ee
and (as first articulated in early searches for realistic string compactifications by Dine and Seiberg \cite{Dine:1985he}) this leads to the problem. 

On one hand, if the leading term is positive then the scalar potential slopes off towards zero in the limit of vanishing string coupling $1/ s \to 0$ and infinite volume $\tau \to \infty$. But this  stationary point of the potential corresponds to 10D flat space and so does not describe what we see around us, and lies beyond the reach of the 4D EFT. Ref.~\cite{Dine:1985he} then argues that if the potential has a non-trivial minimum, as is required to avoid the runaway to infinity, different orders in these expansions must compete with one another ({\it e.g.}~quantum effects must compete with the classical results) signalling the breakdown of the perturbative expansion itself. They concluded that the generic weak-coupling situation is a runaway without a non-trivial minimum. Conversely, if a non-trivial minimum exists then it should generically arise at strong coupling with an extra-dimensional volume of order the string scale: $s \sim \tau \sim 1$. This argument can also be cast in terms of two accidental approximate scale invariances that turn out to be shared by all string vacua, for which $s$ and $\tau$ play the role of pseudo Goldstone modes and 10D flat space corresponds to the scale invariant point (see for instance \cite{Burgess:2020qsc}). 

Of course the key word in this argument is `generic'. Over the years many efforts were made to overcome this general problem and obtain weak couplings and large hierarchies in controlled ways. The solutions usually exploit the few parameters that are not $vevs$ of moduli that can be adjusted to provide non-generic solutions with weak coupling and large volume. These parameters include the curvature of the extra dimensions, non-critical dimensionality, the ranks of the various symmetry groups, or integer flux quantum numbers for antisymmetric tensor fields that thread compact extra dimensions (similar to magnetic flux threading a sphere -- see \cite{Douglas:2006es} for a review). 

In particular IIB string compactifications have been much explored with successful scenarios using a combination of the huge number of possible fluxes and various small non-perturbative effects, leading to two main approaches for stabilizing moduli in IIB vacua. The first of these -- the `KKLT' scenario -- exploits the vast number of fluxes to tune the tree-level superpotential to be exponentially small (so as to compete with small non-perturbative contributions to the superpotential \cite{Giddings:2001yu, KKLT}). The second class -- the `large-volume scenario' or LVS -- instead finds solutions with stabilized moduli by balancing different orders of different expansions, exploiting the fact that generically there are many moduli and many perturbative expansions going on at the same time. In particular non-trivial vacua are found for which the non-perturbative corrections to the superpotential $W_{np}\simeq e^{-a \tau_s}$ for one modulus $\tau_s$ compete with the perturbative corrections in the much larger volume modulus $\cV=\tau^{3/2}$, resulting in minima for which $\tau \simeq e^{a\tau_s}$ and so give exponentially large volumes \cite{Balasubramanian:2005zx, Conlon:2005ki}. For both scenarios the resulting potential is minimized with $V_{\rm min} < 0$ and so additional `uplift' mechanisms are required to obtain flat or de Sitter space, using extra ingredients such as antibranes, T-branes, {\it etc.}~\cite{KKLT, Cicoli:2015ylx,Cicoli:2012fh}. These scenarios have been explored in considerable detail and so far represent the state of the art for moduli stabilization in general and de Sitter string solutions in particular.

In \S\ref{sec:dS} we present an alternative mechanism for moduli stabilization that shares some of the attractive properties of KKLT and LVS pictures, but also evades some of their difficulties. Our proposal differs substantially from both of them by being purely based on perturbative corrections within the corresponding effective field theory (EFT), but doing so in a way consistent with the Dine-Seiberg problem. Our scenario adapts a proposal made for higher-dimensional theories in \cite{AndyCostasnMe} (and further elaborated for 4D supergravity in \cite{YogaDE}). 

The idea is very simple and can be illustrated with a concrete toy model. To this end, consider the perturbative expansion for the low-energy scalar potential $V$ in powers of the volume field $\tau$,
\be
V(\tau)=\sum_{n} A_n(s) \tau^{-n}
\ee
where naively the coefficients $A_n$ depend on all of the other moduli such as $s$. It is tempting to think that the $A_n$ should also be independent of $\tau$, but this need not be true because $A_n$ can depend logarithmically on $\tau$. It is generic that quantum corrections can introduce anomalous scaling into effective interactions, which become logarithmic dependence on ratios of particle masses in a perturbative regime for which a small expansion parameter $\alpha$ exists.  (This expansion parameter might simply be another modulus, like $\alpha\simeq 1/s$.) But in string theory particle masses generically depend on $\tau$ since this field determines the ratio of fundamental scales like the string, Planck and Kaluza-Klein (KK) masses ($M_s$, $M_p$ and $M_\KK$). Consequently any logarithmic dependence on ratios of masses can also imply a logarithmic dependence on $\tau$ (and on any other moduli that appear in mass ratios).  

Because this dependence has its roots in anomalous scaling, renormalization group (RG) techniques can be used to resum leading-log effects ({\it i.e.}~they allow one to work to all orders in the expansion $\alpha \ln\tau$ while still neglecting subdominant terms like $\alpha^2\ln\tau$). Therefore, even though $A_n$ might only be known perturbatively in $\alpha$, RG reasoning gives this expansion to all orders in $\alpha \ln\tau$. The resulting potential can be minimized with respect to $\tau$ without going beyond leading order in the $1/\tau$ expansion, and naturally leads $\ln\tau$ to be fixed at a size $\ln\tau\simeq 1/\alpha$. For weak coupling $\alpha\ll 1$, this stabilizes $\tau$ at exponentially large values (providing a novel explanation for large hierarchies), and the RG allows this to be done without losing perturbative control. This is a key part of why we can evade the Dine-Seiberg conclusion, and suggests the name {\it RG stabilization} (see \cite{Antoniadis:2019rkh} for a related proposal without the RG overlay).

We see that the modulus $\tau$  can easily be exponentially large in RG stabilization (similar to LVS stabilization), and this in turn reinforces the logic of working within the $1/\tau$ expansion. Also like the LVS case, the RG proposal requires generically  at least two expansions: the $1/\tau$ and the $\alpha$ expansions. Because the perturbative corrections modify the supergravity K\"ahler potential (rather than the superpotential), the non-trivial minima we obtain can equally well be de Sitter or anti de Sitter and so a separate uplifting mechanism is not necessary (unlike for KKLT and LVS pictures).

\subsection*{Application to inflationary dynamics}

Where de Sitter solutions exist, can inflationary solutions be far behind? Interest in inflation is driven by its successful description of observed primordial fluctuations in terms of quantum effects amplified by inflationary expansion. The observation that the fluctuation amplitude points to energy scales not too far below the Planck scale has raised hopes that these primordial fluctuations might eventually provide an observational window on very-high-energy physics. 

This has stimulated the development of a great variety of string-inflationary scenarios over the past decades (see \cite{Baumann:2014nda, Cicoli:2011zz} for reviews), for which modulus-stabilization again plays a crucial role. The various inflationary scenarios choose different moduli to be the inflaton, with both the scalar and axionic components of supersymmetric complex scalars playing important roles in different pictures. Many of these proposals are quite promising, with some contenders rising to the top as observations have become more constraining on theoretical models \cite{SIsuccess}. They also tend to share the following three challenges.
\begin{itemize}
\item {\it Modulus stabilization}. A theory of modulus stabilization is always a prerequisite for any microscopic inflationary model. After all, there is no point arranging the scalar potential to be very shallow along a putative inflaton direction in field space if the potential also turns out to be much steeper in other directions; a slowly rolling field prefers to evolve in the steepest direction available. One way to avoid motion in other steeper directions is to arrange for local minima in these directions into which noninflationary moduli can be trapped.  

\item{\it Fragility}. The extreme shallowness of the potential required for slow-roll inflation can easily be overwhelmed by corrections. In particular, classical scalar masses are notoriously sensitive to quantum effects, but having a small slow-roll parameter $\eta=M_p^2\, V''/V \ll 1$ implies the squared-mass of the inflaton, $m^2 \sim V''$, must be much smaller than the Planck-suppressed Hubble scale, $H_\ssI^2 \sim V/M_p^2$. This makes a slow roll hostage to the many corrections to the inflaton mass that are of Hubble size or larger. Many string inflation scenarios are in particular plagued by a specific version of this fragility -- called the `$\eta$ problem' -- that arises when inflation occurs within a 4D supergravity framework (and is discussed in more detail below).

\item{\it Inflation vs SUSY breaking scales}. String models often produce inflation at high scales,  such as the GUT scale $M_{\GUT}\simeq 10^{17}$ GeV, particularly if they are designed to maximize tensor-mode signals \cite{Silverstein:2008sg, Cicoli:2008gp}. However such constructions also tend to give a very large supersymmetry breaking scale. This need not be a serious problem, but makes specific realizations of low-energy supersymmetry breaking difficult if combined with inflation since it creates a tension between the large inflationary scale and the low scale of supersymmetry breaking \cite{Kallosh:2004yh,Conlon:2008cj, Cicoli:2015wja}.
\end{itemize}
In \S\ref{sec:NLSUSYInf} we explore some of the implications of RG modulus stabilization for string-inflationary models, with the encouraging news that it helps resolve all three of these challenges. 

We do so using the specific example of the brane-antibrane inflationary scenario, for which the inflaton is proposed to be the extra-dimensional separation between a mutually attracting brane and antibrane. Separations between supersymmetric BPS branes were proposed as candidates for the inflaton some time ago \cite{Dvali:1998pa}, with slow roll occurring because of the absence of inter-brane forces implied by the BPS condition. The difficulty breaking supersymmetry in these models made it hard to compute the inflaton potential, however, though this was solved by instead exploiting the Coulomb attraction of non-supersymmetric brane-antibrane configurations \cite{Burgess:2001fx,Dvali:2001fw}, though it was realized early on that slow roll remained difficult to achieve in simple geometries because the branes could not be sufficiently separated within the extra dimensions to allow them to experience a weak enough force \cite{Burgess:2001fx}. 

More detailed string constructions required modulus-stabilization techniques \cite{Giddings:2001yu, KKLT}, since only these allow the calculation of the scalar potential for all low-energy moduli. Although it was initially hoped that extra-dimensional warping might potentially resolve the `runway-length' problem \cite{KKLT, KKLMMT}, it turned out that modulus stabilization carried a sting because it also robustly introduced the $\eta$ problem \cite{KKLMMT}, requiring parameters to be tuned in a way that undermined warping's utility. We find the $\eta$ problem is evaded within our stabilization mechanism -- largely because it does not rely on the non-perturbative superpotential as do both KKLT and LVS stabilization -- and so allows the utility of warping for inflation to be resurrected as originally intended.

\subsection*{The inflaton and late-time relaxation}

The advantages of RG stabilization for inflation are not tied to the details of specific string constructions, however. To emphasize this, much of the inflationary discussion of \S\ref{sec:NLSUSYInf} is cast purely in terms of approximate symmetries of the low-energy 4D effective theory. Although the required symmetries are in particular generic to string vacua, phrasing the analysis in terms of low-energy symmetries helps identify what is required (and what is not) in any particular UV completion.  We believe it also undermines the evidence for the swampland hypothesis \cite{Vafa:2005ui, Obied:2018sgi} because it shows how properties of the low-energy theory (like putative difficulty obtaining de Sitter vacua) can be seen as relatively mundane consequences of symmetries rather than indicating any deep failure of EFT methods \cite{Burgess:2020qsc}.

Our inflationary analysis relies on the following two crucial components:
\begin{enumerate}
\item {\it Constrained supersymmetry}: We use the lagrangian for nonlinearly realized 4D supergravity: {\it i.e.}~a supersymmetric gravity sector coupled to a matter sector within which supersymmetry is badly broken ({\it i.e.}~whose superpartners have been integrated out, and so supersymmetry is nonlinearly realized using a goldstino field $G$). In practice this goldstino can be represented using a nilpotent superfield ($G \in X$, with $X^2=0$) \cite{Komargodski:2009rz}. The coupling of such a sector to supergravity can be found in \cite{NilpotentSUGRA}. 

This kind of EFT is of interest in situations where mass splittings within gravitationally coupled supermultiplets are much smaller than splittings in other supermultiplets \cite{LowESugra}, such as often occurs in extra dimensional models \cite{SLED} and string vacua \cite{NilpotentStringVacua, Aparicio:2015psl, Dudas:2019pls} when supersymmetry is badly broken on a brane. It also plausibly applies to Standard Model phenomenology within models that are supersymmetric in the UV, given the evidence for the absence of weak-scale supersymmetry \cite{PDG}.

\item {\it Accidental approximate scale invariance}. We take our representative modulus to be a dilaton $\tau \in T$ that sits within a supermultiplet $T$ and acts as the pseudo-Goldstone boson for an approximate accidental scale invariances. Such scaling symmetries have long been known to be present for specific string vacua \cite{Witten:1985bz, Burgess:1985zz}, although we now know them to be generic for extra-dimensional supergravity \cite{XDsugraScaling}, which in turn very robustly inherit them from string theory \cite{Burgess:2020qsc}.
\end{enumerate}

To these we also add the inflaton field, $\phi$, assuming it to be part of the sector that badly breaks supersymmetry (and so to realize supersymmetry only nonlinearly). An attentive reader might be struck by the similarity between this list of ingredients and those used in ref.~\cite{YogaDE}, for which the only difference is that there the field $\phi$ instead enters as a `relaxation' field whose presence dynamically helps suppress the size of the scalar potential at its minimum. 

The appearance of $\phi$ in \cite{YogaDE} is at first sight a bit jarring, since it appears to arise in a bespoke way with nothing to do with  any other physics. We argue here that one way to understand the presence of $\phi$ is as an inflaton: its very definition requires it to interpolate between a potential that is dominated by a large positive vacuum energy and one where the potential energy is small. (The only surprise in  \cite{YogaDE} is in just how small the potential at this minimum turns out to be.) All that is needed to make early-universe evolution of $\phi$ into an inflationary mechanism is a reason why this transition should happen slowly (and showing how this occurs is the baton we take up here).

Furthermore, the brane-antibrane system lends itself perfectly to this scenario. First, it does so because it is well known that the supersymmetry breaking of an antibrane is captured precisely by the nilpotent superfield $X$. Second, the brane-antibrane Coulomb interaction is also easily captured by a superpotential that couples the inter-brane distance $\phi$ to the nilpotent superfield $X$. Brane-antibrane attraction thereby provides a natural UV interpretation for the relaxation/inflaton field $\phi$.

In what follows we describe in detail in \S\ref{sec:NLSUSYInf} how RG stabilization resurrects warped brane-antibrane inflation, after first describing the RG stabilization mechanism in \S\ref{sec:dS}. \S\ref{sec:dS} also includes discussions of how to realize the RG mechanism in IIB string theory; a determination of the relevant scales, like gravitino mass and soft-breaking terms for the matter sector and their implications for the size of the volume modulus $\tau$; the relevance of RG stabilization to the Dine-Seiberg problem; and why runaway regions can be sensibly addressed using EFT methods (including comparisons with calculations in other areas of physics).

\section{de Sitter vacua}
\label{sec:dS}

In this section we explore the simplest stabilization example. Our goal is to illustrate how standard renormalization-group methods can allow modulus stabilization without losing perturbative control. We also use this example to show how combining this stabilization mechanism with accidental low-energy symmetries (in particular accidental, approximate scale invariance) leads to a novel kind of spontaneous supersymmetry breaking that generates de Sitter solutions with large hierarchies driven by the exponentially large value of the stabilized moduli. We first present the mechanism in a stripped-down model and then show how it naturally embeds into low-energy string vacua.

\subsection{Accidental symmetries and dilaton dynamics}

Consider a general low-energy 4D effective theory that is both supersymmetric and enjoys an accidental approximate scaling symmetry (of the form argued in \cite{Burgess:2020qsc} to be generic in low-energy string vacua and in \cite{XDsugraScaling} to be generic to higher-dimensional supergravities more generally). The minimal such a model involves the gravity supermultiplet and the chiral superfield $T \ni \{\cT, \xi \}$ that contains a complex scalar $T = \frac12(\tau + i \mfa)$ whose real part ($\tau$) is the dilaton required by the approximate scale invariance. Because the EFT is a 4D supergravity it is determined at the two-derivative level in terms of standard supergravity ingredients: by specifying how the K\"ahler potential $K$, the superpotential $W$ and (should gauge multiplets also be present) the gauge kinetic function $\mff_{\alpha\beta}$ depend on the one chiral superfield $T$.  

We take $W = w_0$ to be independent of $T$ (as can be enforced with the axionic symmetry under shifts of $\mfa$). Accidental approximate scale invariance is implemented by demanding that $e^{-K/3}$ arises as a series in powers of $1/\tau$, as in:
\be\label{RelaxK}
  e^{-K/3} = \tau - k + \frac{h}{\tau} + \cdots  \,,
\ee
where the ellipses denote higher orders in $1/\tau$. This ensures that the lagrangian density comes as a series of terms, $\cL = \sum_n \cL_n$, each of which scales homogeneously, $\cL_n \to \lambda^{p_n} \cL_n$ for some $p_n$, when $g_{\mu\nu}$ and $\tau$ are scaled by powers of the constant parameter $\lambda$. This kind of expansion ensures that semiclassical methods arise as expansions in powers of $1/\tau$ and so are good approximations in the regime $\tau \gg 1$ --- and so this is where we seek our minima once we compute a potential $V(\tau)$. It represents a scale invariance because these rescalings of $\tau$ and the metric are symmetries of the classical field equations to leading order in $1/\tau$. 

String theorists will recognize this system: in Type IIB string compactifications the role of the field $\tau$ is played by the one always-present K\"ahler modulus: the extra-dimensional volume (in string units), $\cV \propto \tau^{3/2}$. In a string context the reliance of semiclassical arguments on large $\tau$ expresses how semiclassical supergravities provide reliable EFTs for string vacua only for geometries that are much larger than the string scale.

Crucially, although powers of $\tau$ are explicit in \pref{RelaxK}, in general quantum effects complicate the scaling properties of subdominant terms in the lagrangian. We return below to why this is so, but just record now that it allows the functions $k$, $h$ to be rational functions of logarithms of $\tau$: $k = k(\ln\tau)$, $h = h(\ln\tau)$ {\it etc.}. 

The lagrangian obtained with these choices for $K$ and $W$ have the familiar supergravity form, with the Einstein-frame kinetic term for the bosons given (in Planck units) by
\be \label{TkinL0}
  - \frac{\cL_{\rm kin}}{\sqrt{-g}} = \frac12 \, \cR + K_{\ssT \ol\ssT} \, \partial_\mu \cT\, \partial^\mu \ol \cT \simeq \frac12 \, \cR + \left( \frac{3 }{\tau^2} + \cdots \right) \partial_\mu \cT\, \partial^\mu \ol \cT   \,,
\ee
where (as usual) $\cR$ denotes the Ricci scalar built from $g_{\mu\nu}$ and subscripts on functions like $K$ and $k$ denote differentiation with respect to the fields: {\it e.g.}~$K_{\ssT\ol\ssT} = \partial_\ssT \partial_{\ol\ssT} K$. 

The scalar potential is similarly given by
\be \label{VFdef2}
  V = e^{K} \Bigl[ K^{\ol \ssT \ssT}\, \ol{D_\ssT W} D_\ssT W - 3|W|^2 \Bigr]   \,.
\ee
where $K^{\ol \ssT \ssT} = 1/K_{\ssT \ol \ssT}$ and
\be \label{DTWform}
   D_\ssT W = W_\ssT + K_\ssT W \simeq \left( - \frac{3}{\tau} + \cdots \right) w_0 \,.
\ee
The last equality uses \pref{RelaxK} for $K$. The leading parts of the scalar potential then are
\be \label{VFtauexp}
  V  \simeq  - \frac{3 \, k_{\ssT\ol\ssT}}{\cP^2} \,  |w_0|^2+ \cdots = \frac{3 \, (k' - k'') }{\tau^4} \,  |w_0|^2+ \cO(\tau^{-5}) \,,
\ee
where $\cP := e^{-K/3} = \tau - k + \cdots$ and primes denote differentiation with respect to $x = \ln \tau$. Notice that expression \pref{VFtauexp} vanishes whenever $k$ is independent of $T$, as it must do on general grounds because \pref{RelaxK} becomes a no-scale model \cite{NoScale} in the limit that $h$ (and higher terms) vanish and $k$ is $T$-independent. For later purposes recall also that $k_{\ssT\ol\ssT}$ can have either sign since it does not control the sign of the kinetic energy for $\cT$ in \pref{TkinL0}. Contributions involving $h$ and other subdominant terms in \pref{RelaxK} first arise at order $\cO(\tau^{-5})$. 
 
\subsection{Controlled perturbative stabilization}
\label{ssec:taustab}

Eq.~\pref{VFtauexp} reveals that the leading contribution to the potential for large $\tau$ has the form
\be \label{VFBasicForm}
  V(\tau)  \simeq   \frac{U(\ln\tau)}{\tau^4}  \,,
\ee
with $U(\ln\tau) = - 3 \tau^2 k_{\ssT\ol\ssT} |w_0|^2 = 3(k'-k'')|w_0|^2$. The minima of \pref{VFBasicForm} depend on the functional form of $U$ and so requires more information about how $k$ acquires its dependence on $\ln\tau$.

To this end, following the ideas of \cite{AndyCostasnMe} and \cite{YogaDE}, we imagine that $k$ acquires its dependence on $\ln\tau$ through the running of some dimensionless coupling $\alpha_g$, due to a perturbative expansion of the form
\be \label{Kvsalpha}
   k \simeq k_0 + k_1 \, \alpha_g + \frac{k_2}{2}\, \alpha_g^2 + \cdots 
\ee
with a dimensionless coupling $\alpha_g \ll 1$. In general the running of a dimensionless coupling like $\alpha_g$ introduces logarithms of mass ratios, such as when its renormalization-group evolution is integrated to give
\be \label{alphavsmu}
   \frac{1}{\alpha_g(m_1)} =  \frac{1}{\alpha_g(m_2)}- b_1 \ln \left( \frac{m_1}{m_2} \right) \,.
\ee
The main observation is that this can become a dependence\footnote{More precisely masses actually develop a dependence on $\cP$ rather than just $\tau$ because they typically acquire their leading dependence on $\tau$ through powers of the Weyl rescaling factor $e^{-K/3} = \cP = \tau - k + \cdots$. This makes $k$ a function of $\ln\cP$ rather than $\ln\tau$ in the discussions to follow; a distinction that often does not matter, but plays an important role when discussing the $\eta$ problem for the inflationary scenarios of \S\ref{ssec:WBABinf}.} on $\ln\tau$ if there are multiple fields coupling to this interaction whose masses\footnote{Notice that it is only ratios of physical masses that matter here and {\it not} ratios of masses to the RG running parameter $\mu$. This is because any $\tau$-dependence associated with $\mu$ ultimately cancels from physical observables for the same reason that all $\mu$-dependence also cancels.} depend differently on $\tau$. 

Such interactions do plausibly arise in string compactifications. For instance in IIB compactifications particles localized on D3 and D7 branes have masses that depend differently on the volume modulus and when such branes intersect they can both couple to light open-string 4D gauge fields (whose gauge coupling could be the $\alpha_g$ considered here).

In such a situation \pref{Kvsalpha} predicts a logarithmic $\tau$ dependence for $k$ that emerges through the $\tau$-dependence of $\alpha_g$, which in turn can be expressed through a renormalization-group evolution like
\be \label{betafunctiontau}
\tau \,\frac{\exd \alpha_g}{\exd \tau} = \beta(\alpha_g) =  b_1 \alpha_g^2 + b_2\, \alpha_g^3 + \cdots \,.
\ee
For $\alpha_g$ small enough to neglect all but the leading term in $\beta$ this has solution 
\be \label{alphavstauRGNAPP}
   \alpha_g(\tau) = \frac{\alpha_{g0}}{1 - b_1 \, \alpha_{g0}\,  \ln \tau} \,,
\ee
for some integration constant $\alpha_{g0} = \alpha_g(\tau=1)$. For the present purposes what is important about the $\ln\tau$ dependence given in \pref{alphavstauRGNAPP} is that its derivation neglects only additional powers of $\alpha_g$ in \pref{betafunctiontau}. Consequently for large $\tau$ it remains valid to all orders in $\alpha_g \ln\tau$ while dropping contributions of order $\alpha_g^2 \ln \tau$. It is this renormalization-group resummation that ultimately allows us to trust minima of the potential that occur in the regime $\ln \tau \sim 1/\alpha_g$. 

Now comes the main point. Using \pref{Kvsalpha} and \pref{betafunctiontau} to evaluate the $T$-derivatives of $k$ then gives $k' = (k_1 + k_2 \, \alpha_g + \cdots) \beta(\alpha_g)$ and similarly for $k''$, and using these in \pref{VFBasicForm} then leads to the expression  
\be \label{Uvsalpha}
   U \simeq  U_1 \, \alpha_g^2 - U_2 \, \alpha_g^3 +  U_3 \, \alpha_g^4 + \cdots \,,
\ee
where $U_1 = 3 k_1 b_1|w_0|^2$ and so on. Furthermore, the Dine-Seiberg argument leads one to expect that any minima $\tau = \tau_0$ of this potential generically occur in the regime where $\alpha(\tau_0) \sim \cO(1)$. But if stabilization of other moduli make $\alpha_{g0}$ small, then inspection of \pref{alphavstauRGNAPP} shows that $\tau_0$ must be very large because $\alpha_{g0} \ln \tau_0 \simeq \cO(1)$.

\begin{figure}[t]
\begin{center}
\includegraphics[width=120mm,height=60mm]{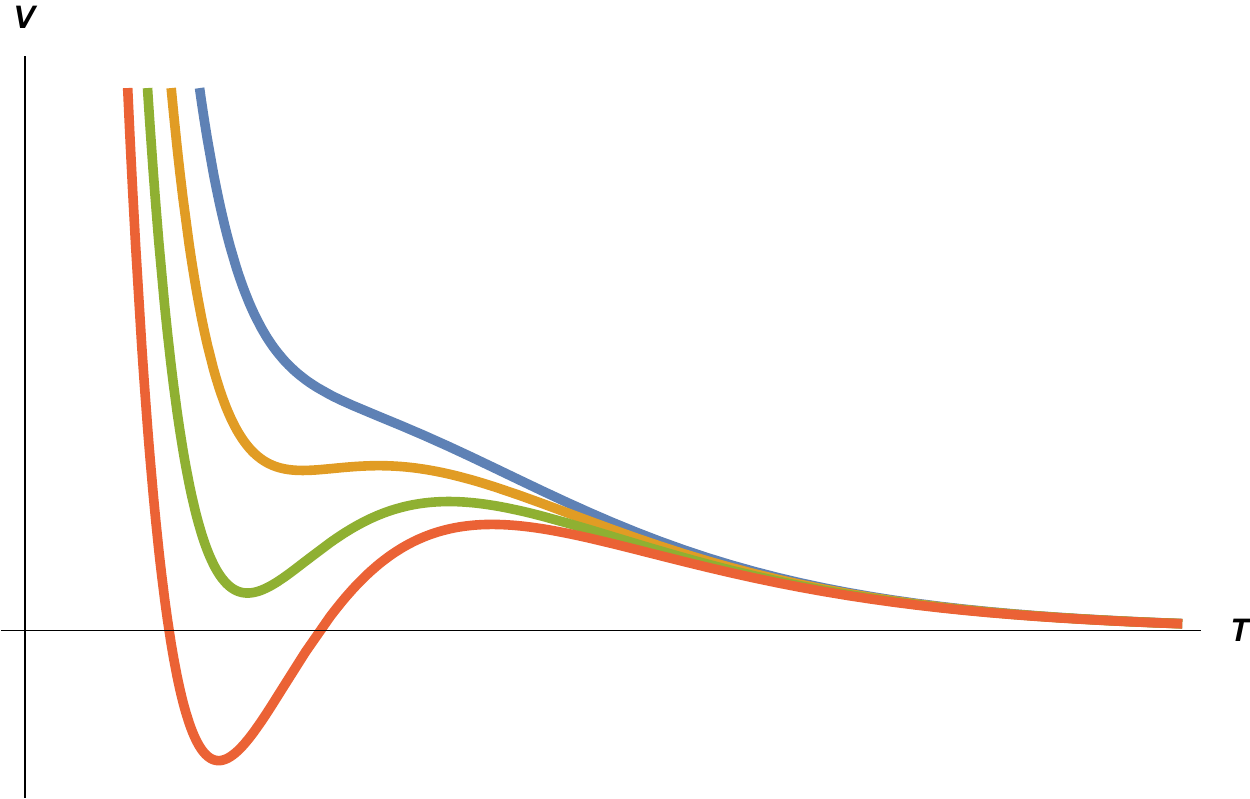} 
\caption{A plot of $V$ vs $\tau$ for the scalar potential $V=U(\ln\tau)/\tau^4$, revealing a de Sitter or anti-de Sitter minimum separated from a runaway by a local maximum. The plots are obtained using the representative values $k_1/k_3 = 0.01$ and $k_2/k_3 = - 0.133$ (arbitrary scale). The main text describes the precise parameter range required to get de Sitter rather than anti-de Sitter or a runaway.} \label{Fig:Valpha} 
\end{center}
\end{figure}

This general argument can be made explicit purely using perturbative methods if we arrange that the coefficients $U_1$, $U_2$ and $U_3$ appearing in the potential \pref{VFBasicForm} with $U$ given by \pref{Uvsalpha} are all positive and satisfy the mild hierarchy  
\be\label{coeffhier}
   \left| \frac{U_1}{U_2} \right| \sim \left| \frac{U_2}{U_3} \right| \sim \cO(\epsilon)
\ee
for some smallish $\epsilon \ll 1$. Such a hierarchy allows solutions to $\left. \partial V/\partial \tau \right|_{\tau_0} = 0$ for $\alpha_0 \sim \cO(\epsilon)$ and so 
\be
   b_1 \ln\tau_0 = \alpha_{g0}^{-1} - \epsilon^{-1}
\ee
can easily be order $1/\epsilon$ if $\epsilon \ll \alpha_{g0}$ and $b_1 < 0$. For $\epsilon \lsim 1/10$ the value predicted for $\tau_0$ can be enormous $\tau_0\sim e^{1/\epsilon}$, justifying the validity of the $1/\tau$ expansion {\it ex post facto}. As is easy to check, when $9 \, U_2^2 > 32\, U_1 U_3$ the potential has a local minimum at $\tau_0$ that is separated from the runaway to $\tau \to \infty$ by a local maximum at $\tau_1 > \tau_0$ (see Fig.~\ref{Fig:Valpha}).

The value of the potential at this minimum is positive if $U_2^2 < 4\, U_1 U_3$ and negative otherwise. Although \pref{Uvsalpha} and \pref{coeffhier} might naively lead one to expect $U(\tau_0) \sim \cO(\epsilon^4)$ when $U_3 \sim \cO(1)$, it happens that the condition $V'(\tau_0) = 0$ ensures that this leading contribution cancels, making the result at the minimum instead $U(\tau_0) \sim \cO(\epsilon^5)$. As a result both $V(\tau_0)$ and $\left. \tau^2 (\partial^2 V/\partial\tau^2)\right|_{\tau_0}$ are $\cO(\epsilon^5|w_0|^2/\tau_0^4)$, and this can be extremely small given that $\tau_0$ can be an exponential of $1/\epsilon$. Ref.~\cite{YogaDE} explores some of the implications if this suppression were to explain the size of the present-day Dark Energy density. 

Because $U(\tau_0)$ can have either sign both de Sitter and anti-de Sitter solutions can be generated in this way depending on the values of the coefficients $U_0$, $U_1$ and $U_2$. Both signs are allowed because \pref{DTWform} shows that supersymmetry is broken for any finite $\tau$. It breaks because the auxiliary field $F^\ssT$ for the $T$ supermultiplet is nonzero, since $w_0\neq 0$,  even though $W_\ssT$ vanishes. Its size is instead controlled by the Planck suppressed term $K_\ssT W/M_p^2 \in D_\ssT W$. This type of supersymmetry breaking is common in no-scale models and is responsible for many of the unusual properties encountered in \cite{YogaDE}. This source of supersymmetry breaking is easily missed in global supersymmetry because it disappears in the $M_p \to \infty$ limit.

\subsection{Type IIB string theory realization}

We next expand on how the above mechanism arises in the low-energy limit of Type IIB string vacua. One purpose in doing so is to identify the scales to which this stabilization mechanism points. Another purpose is to see how such an explicitly perturbative mechanism evades the well-known challenges posed by the Dine-Seiberg problem \cite{Dine:1985he}. We discuss each of these issues after first making the connection to IIB vacua more explicit.

The massless bosonic fields in the 10D supergravity relevant to Type IIB vacua below the string scale are 
\be
\tilde g_{\ssM\ssN} ,\quad  \cS=s - i C, \quad G_{(3)}=H_{(3)}+i\cS F_{(3)}, \quad \tilde F_{(5)}=\exd C_{(4)}+\frac12\, C_{(2)} \wedge H_{(3)} + \frac12 \, B_{(2)} \wedge F_{(3)}
\ee
where a subscript $(p)$ indicates that the corresponding field is a $p$-form, $s = e^{-\hat\phi}$ is the 10D dilaton\footnote{The hat on $\hat\phi$ distinguishes the string dilaton from the inflaton field $\phi$ used everywhere else in this paper.} that controls the local string coupling and $C$ is an axionic scalar while $H_{(3)}=\exd B_{(2)}$ and $F_{(3)}=\exd C_{(2)}$ are field strengths for 2-form gauge potentials. At the two-derivative level the action for these fields takes the schematic form
\be
\label{TypeIIBBulk}
 S_{\rm bulk} = \int \exd^{10}x \, \sqrt{-\tilde g} \; \left\{\tilde R - \frac{\lvert \partial \cS \rvert^2}{(\mathrm{Re}\,\cS)^2} - \frac{\lvert G_{(3)} \rvert^2}{\mathrm{Re}\, \cS} - \tilde{F}_{(5)}^2 \right\}  + \int \frac{1}{\mathrm{Re} \,S}\, C_{(4)} \wedge G_{(3)} \wedge \ol{G}_{(3)}   \ ,
\ee
This action has two accidental symmetries that are important for our present purposes:
\begin{itemize}
\item 
An $SL(2,\mathbb{R})$ symmetry under which 
\be\label{SL2Rdef}
  S \to \frac{a \,\cS - i b}{ic\, \cS + d} \quad \hbox{and} \quad 
  G_{(3)} \to \frac{G_{(3)}}{i c\, \cS + d}  \,,
\ee
where $ad-bc =1$. Note that the special case $b=c=0$ and $a = 1/d$ reduces to a classical scaling symmetry
\be
\label{IIBscalings1}
 \tilde g_{\ssM\ssN} \rightarrow \tilde g_{\ssM\ssN} \ , \qquad \cS \rightarrow  a^2 \cS\ , \qquad G_{(3)} \rightarrow a G_{(3)} \,, \qquad \tilde F_{(5)} \rightarrow  \tilde F_{(5)} \,.
\ee
\item
An approximate accidental scale invariance
\be
\label{IIBscalings}
 \tilde g_{\ssM\ssN} \rightarrow \lambda \tilde g_{\ssM\ssN} \ , \qquad \cS \rightarrow  \cS\ , \qquad B_{(2)} \rightarrow \lambda B_{(2)} \,, \qquad C_{(2)} \rightarrow \lambda C_{(2)} \,.\qquad C_{(4)} \rightarrow \lambda^2  C_{(4)} \,.
\ee
under which the tree level action scales as $ S_{bulk} \rightarrow \lambda^4 S_{bulk}$. Upon compactification to four dimensions the non-trivial scaling of the 10D metric implies an overall scaling of the volume modulus $\cV\rightarrow \lambda^3 \cV$.
\end{itemize}

These two approximate symmetries are accidental in the sense that they are broken by $\alpha'$ and loop corrections to the effective action. Indeed, how terms scale under these two transformations can be used to identify how the 10D action depends on these two expansions \cite{Burgess:2020qsc}. For the 4D theory, the $\alpha'$ expansion becomes an expansion in inverse powers of the volume $\cV:=\tau^{3/2}$ while the string-loop expansion is in powers of $(\hbox{Re} \,\cS)^{-1} = e^{\hat\phi}$. 

Both scaling symmetries are spontaneously broken inasmuch as neither leaves generic background fields unchanged and the volume modulus and the string dilaton can be regarded as their pseudo-Goldstone dilaton modes. From this point of view 10D flat space is special inasmuch as it leaves a scale invariance unbroken because \pref{IIBscalings} does not act on $\cS$ and scale transformations of the flat metric can be compensated by a diffeomorphism. 10D flat space corresponds in 4D to $\cV \to \infty$ and $s \to \infty$ and the scale-invariance of this point anchors the asymptotic value of the 4D scalar potential to zero.

\subsubsection{IIB modulus stabilization}

String theory famously has no parameters, but if so what are the choices that lead to differently shaped compactifications? For IIB Calabi-Yau orientifold compactifications the choices made are the quantized fluxes of the three-form fields whose presence and stress-energy stabilizes the complex structure moduli $U$ and string dilaton $\cS$, as pioneered in \cite{Giddings:2001yu}. For supersymmetric flux configurations these moduli are fixed in the 4D effective description by the supersymmetric conditions $D_\ssS W = D_\ssU W =0$. 

The choice of higher-dimensional fluxes shows up in the low-energy 4D theory in several ways. First, $(0,3)$ fluxes induce a non-trivial superpotential $w_0 \in W$ that is generically order unity (in Planck units) but can be arranged to take larger -- or extremely small \cite{Demirtas:2021nlu} -- values. Second, since fluxes fix the string dilaton field $\cS$  they provide a `discretuum' of possible values for the string coupling constant $g_s\sim s^{-1}$ in the 4D theory. Third, fluxes can fix the complex structure moduli in such a way that the corresponding three-cycles in the extra-dimensional geometry become long throats along which 4D geometries are naturally warped with warp factor $e^{\cA}\sim e^{8\pi K/g_sM}$, where $K,M$ are integers. The three quantities $w_0$, $g_s$ and $e^{\cA}$ play important roles defining the different scales that arise within the 4D theory.

The Calabi-Yau space's K\"ahler moduli are not similarly fixed by these fluxes and so their potential is naturally explored within the 4D theory. The simplest case arises for Calabi-Yau orientifiolds that have the fewest possible K\"ahler moduli: the single complex modulus $T= \frac12(\tau+i \mfa)$ whose real part describes the overall volume $\cV \propto \tau^{3/2}$ of the Calabi-Yau and whose imaginary part $\mfa$ is an axionic partner. The shift symmetry for this axion $\mfa\rightarrow \mfa+c$ can forbid\footnote{Whether it does or not depends on whether the corresponding symmetry has an anomaly. If so $W$ can depend exponentially on $T$. $T$-dependent corrections to $W$ that are perturbative in $1/T$ are forbidden by the supersymmetric non-renormalization theorems \cite{NRTheorems, Witten:1985bz, Burgess:1985zz}.} its appearance in the superpotential $W$. The leading expression for the K\"ahler potential for $T$ is well known to be of the no-scale type 
\be \label{CYK}
  K(T,\ol T) = -2 \ln \cV = -3 \ln \tau \,,
\ee
as can be derived either from explicit dimensional reduction or using the transformation properties of the 4D action under the approximate accidental scaling symmetries \pref{IIBscalings1} and \pref{IIBscalings}. 

As discussed earlier, the condition $W_\ssT = 0$ together with the no-scale identity $K^{\ol\ssA\ssB} K_{\ol\ssA} K_\ssB = 3$ satisfied by the K\"ahler potential \pref{CYK} ensures the scalar potential \pref{VFdef2} is independent of $\tau$ and this precisely reproduces the microscopic statement that K\"ahler moduli are not fixed in the underlying flux construction at leading order in string coupling and $\alpha'$. K\"ahler modulus stabilization proceeds because higher-order corrections lift this flatness and so can stabilize fields like $T$. At present the main approaches to modulus stabilization drive this stabilization by introducing a $T$-dependent contribution to the superpotential, which can arise nonperturbatively in $1/\tau$ through contributions of the form $\delta W=W_{np}\propto e^{-\xi \, T}$ for some $\xi$. Introducing $T$-dependence to $W$ lifts the flatness of the no-scale potential, and can be consistent with the underlying $1/\tau$ expansion either if $w_0$ happens to be extremely small \cite{KKLT} or by considering multiple K\"ahler moduli, $\tau_v$ and $\tau_s$ and having $\tau_v \sim e^{\,\xi \tau_s}$ so that powers of $1/\tau_v$ can compete with $\delta W \propto e^{-\xi \, T_s}$ \cite{Balasubramanian:2005zx, Conlon:2005ki}.  

We instead here do not introduce a $T$-dependence to $W$ at all, and considering only perturbative corrections to $K$ (that would in any case normally dominate over non-perturbative effects). Denoting $s = \hbox{Re}\, \cS = e^{-\phi}$, in general \cite{Burgess:2020qsc} perturbative corrections to $K$ in powers of $1/s$ and $1/\tau$ can be written as
\be \label{KcorrIIB}
e^{-K/3} = s^{1/3} \tau\sum_{nmr} \cA_{nmr} \left( \frac{1}{s} \right)^{n} \left( \frac{s}{\tau} \right)^{(m+r)/2}  \,, 
\ee
with $n$ counting string loops and the $\alpha'$ expansion receives contributions from $r$ powers of extra-dimensional curvature and $m+1$ powers of 3-form flux $G_{(3)}$. The coefficients $\cA_{nmr}$ here are to be regarded as functions of all other moduli\footnote{A crucial difference in our approach is to consider that the coefficients $\cA_{nmr}$ can have a $\ln\tau$ dependence.}, but the powers of $s$ and $\tau$ associated with the string-loop and $\alpha'$ expansions are explicit. Tracking only the volume dependence then shows that the K\"ahler potential can be written as the following expansion in powers of $1/\tau$ 
\be\label{Kexpansion}
  K(T,\ol T)=-3   \ln \cP,\quad \hbox{with} \quad \cP(\tau) = \tau\left[ 1 - \frac{k}{\tau} + \frac{h}{\tau^{3/2}} + \cO\left(\frac{1}{\tau^2}\right) \right] \,.
\ee  
This has a form very similar to \pref{RelaxK}, differing\footnote{Notice that in principle \pref{KcorrIIB} allows a contribution with $m+r=1$ that, if present, would change \pref{Kexpansion} to $\cP=\tau(1+g/\sqrt{\tau}-k/\tau+\cdots)$. This in turn would lead to a leading correction to the scalar potential of order $\delta V\sim \cO(\tau^{-7/2})$ that would dominate the contributions we consider here. However no known Calabi Yau produces these terms and it has recently been shown \cite{Cicoli:2021rub} that such corrections are generally absent at least to leading order in string loops. Including $m+r=1$ contributions still allows our stabilization mechanism, but the logarithmic terms in $K$ would arise at order $\tau^{-7/2}$ rather than order $\tau^{-4}$ as we use here.} only by being an expansion in powers of $\tau^{-1/2} \propto \cV^{-1/3}$ rather than $\tau^{-1} \propto \cV^{-2/3}$. 

As discussed above, the corrections in \pref{Kexpansion} generically lift the no-scale flat direction, but when the coefficients $k$ and $h$ are $\tau$-independent the leading contribution to $V$ comes from $h$ and arises at order $\delta V\sim \cO(\tau^{-9/2})$, corresponding to what are $(\alpha')^3$ corrections in the underlying string construction. The persistence of the potential's flatness in the presence of a $T$-independent $k$ is known as the compactification's extended no-scale property, and follows because the first two terms of \pref{Kexpansion} still satisfy the no-scale identity $K^{\ol\ssA\ssB} K_{\ol\ssA} K_\ssB = 3$.  

The new ingredient here is not to try to stabilize $\tau$ by balancing different powers of $1/\tau$ in the expansion of $V$ that follow from \pref{Kexpansion}, but instead to recognize that the coefficients $k$ and $h$ can generically contain a $\ln \tau$ dependence  $k=k(\ln\tau)$. This gives the leading contribution to the scalar potential as in \pref{VFBasicForm}, and the minimum is instead obtained by balancing different powers of $\alpha_g \ln \tau$ against one another. As discussed earlier this has the advantage that this balancing can be done without undermining either the $1/\tau$ or $\alpha_g$ expansions. 

We see in this way that the stabilization scenario proposed in \S\ref{ssec:taustab} can apply directly to IIB string theory. It easily gives exponentially large volumes, similar to the large-volume scenario (LVS), and like the LVS requires a second expansion modulus.\footnote{As mentioned earlier, exponentially large volumes arise in LVS through the introduction of a second K\"ahler `blow-up' modulus $\tau_s$ that appears exponentially in the superpotential, with the potential minimized when powers of $1/\cV$ balance against this non-perturbative contribution to $W$ (which occurs when $\cV \simeq e^{a\tau_s}$). It is the coupling $\alpha_g$ itself that would be the required second modulus in the approach we follow here (and so need not be a K\"ahler modulus, such as if it is the inverse string dilaton itself).} Here, naturally, the dilaton $s$ can be fixed by fluxes to give small enough coupling (and so $1/s$ can play the role of $\alpha_g$ in our analysis). This is similar to LVS. But, unlike LVS models, the mechanism presented here does not need to add new uplifting mechanisms to obtain de Sitter space.\footnote{Of course obtaining anti-de Sitter space at tree level need not mean a solution fails to describe our universe because quantum corrections to $V$ often dominate the classical prediction and so can be the source of positive energy that allows de Sitter solutions (for a recent suggestion along these lines see \cite{deAlwis:2021zab}).} The scenario described here also extends straightforwardly to the more general case with more than one K\"ahler modulus (ref.~\cite{YogaDE} explores some multi-modulus examples) inasmuch as the volume is stabilized by leading order contributions of order $\tau^{-4}$ whereas the smaller `fibre' moduli can be stabilized using the next order contributions of order $\tau^{-9/2}$ (such as is done in LVS constructions, and exploited for inflationary purposes in \cite{Cicoli:2008gp}).  

Whether $\ln\tau$-dependence actually appears in $k$ in specific string constructions is of course model dependent. For gauge interactions such log dependence requires the existence of more than one type of matter field for which the masses scale differently with $\tau$ since the logs are sensitive only to ratios of masses. As mentioned earlier, in IIB models this is the case whenever there are chiral states charged under both D3 and D7 gauge groups. A general study of concrete models for which these logs are present is beyond the scope of this article, but we refer the reader to recent discussions on the appearance of logs in IIB EFTs \cite{Conlon:2010ji, Grimm:2014xva, Weissenbacher:2019bfb, Weissenbacher:2020cyf,Klaewer:2020lfg, Antoniadis:2019rkh}. In particular \cite{Antoniadis:2019rkh} computes amplitudes that lead to logarithmic corrections to the K\"ahler potential of order $\delta K = \cO(\ln\cV/\cV)$ and use them as a mechanism to stabilize moduli (although without the renormalization-group resummation used here).

\subsubsection{Scales and Soft terms}

We have seen the value of $\tau$ can be fixed at a wide range of exponentially large values using only a relatively small range of parameters $k_i$, but precisely how big should we like $\tau = \tau_0$ to be at the minimum?  There are two classes of regimes that are natural to consider.

\subsubsection*{Yoga models}

The ambitious point of view asks $\tau_0$ to be large enough that $V_{\rm min} \propto \tau_0^{-4}$ can be as small as the observed dark energy density. This requires $\tau_0 \gsim 10^{26}$ and is the regime explored in some detail in \cite{YogaDE}. For $\tau_0$ this large the mass of the $\tau$ field and its axionic partner are light enough to be cosmologically active in the recent universe. Although one might imagine such light scalars to be ruled out by solar-system and cosmological tests of gravity, ref.~\cite{YogaDE} shows that they are surprisingly hard to constrain, partly due to the appearance of surprising new mechanisms for screening \cite{ADScreening}. 

For the present purposes the main problem with choosing $\tau_0$ this large is that the 4D theory near this minimum requires a UV completion at scales of order $M_p/\tau_0$ because this is where the axion decay constant lies. This in itself need not be a problem because this occurs at eV scales when $\tau_0 \sim 10^{26}$ and so the required UV physics could plausibly be extra-dimensional. Whether this kind of a picture is viable then depends on precisely how $\tau$ arises in the UV completion, but if it does so as a volume modulus along the lines described here then there is a problem. 

The problem arises because the string and Kaluza-Klein scales are related to $\tau$ by
\be \label{MsMKK}
M_s=\frac{M_p}{\cV^{1/2}}\sim \frac{M_p}{\tau^{3/4}} \quad \hbox{and} \quad
M_\KK \sim \frac{M_s}{\cV^{1/6}} = \frac{M_p}{\cV^{2/3}} \sim \frac{M_p}{\tau} \,.
\ee
Requiring $M_\KK \gsim 10$ TeV near the minimum implies $\tau_0 \lsim 10^{14}$, but this bound can be model-dependent because it can be evaded by having some extra dimensions be smaller than others. More robust is the condition $M_s \gsim 10$ TeV, which implies $\tau_0 \lsim 10^{20}$. Because our focus here is on string embeddings we do not pursue values of $\tau_0$ larger than this any further. See \cite{YogaDE} for more in-depth discussion of these issues.

\subsubsection*{SUSY Breaking at TeV and higher scales}

The alternative point of view is to ignore (as most do) the cosmological constant problem and ask how other scales depend on $\tau$. In this case having $V_{\rm min} \propto \tau_0^{-4}$ still suppresses its present-day value relative to other approaches, particularly when $\tau_0$ takes its largest allowed values. In this case the main constraint on the present-day value of $\tau$ actually comes from demanding the field $\tau$ to be heavy enough to avoid the cosmological-modulus problem. This problem is a general constraint on gravitationally coupled relics and requires $\tau$ to be heavy enough to decay before nucleosynthesis, so as not to destroy its successes, which for gravitational-strength decays requires $m_\tau \gsim 30$ TeV  \cite{Coughlan:1983ci}. 

Re-introducing factors of $M_p$, the mass of $\tau$ and of the gravitino are related by
\be
m_\tau = \left(\frac{\tau^2}{M_p^2}\frac{\partial^2 V}{\partial \tau ^2} \right)^{1/2} \sim \frac{\epsilon^{5/2} |w_0|}{\tau^2 M_p^2} \sim \frac{\epsilon^{5/2} m_{3/2}}{\tau^{1/2}}\quad \hbox{where} \quad m_{3/2} \sim \frac{|w_0|}{\tau^{3/2} M_p^2}  \,.
\ee
In these expressions we take $\epsilon \sim \cO(1/10)$ since $\tau_0\sim e^{1/\epsilon}$, but even once this is done the implications for $\tau_0$ of the condition $m_\tau(\tau_0) \gsim 30$ TeV depends on the value of $|w_0|$. We choose two representative benchmarks: $|w_0| \sim M_p^3$ (as is most commonly found in string compactifications) or $|w_0| \sim M_p^3\, \tau_0^{1/2}$ (which is the upper limit on what is possible for a 4D supergravity EFT, since for larger $w_0$ the gravitino mass becomes larger than the Kaluza-Klein scale given in \pref{MsMKK} \cite{Cicoli:2013swa, YogaDE}). Choosing $m_\tau \sim 30$ TeV for each of these cases implies
\bea
   &&\tau_0 \sim 10^6 \,, \quad m_{3/2} \sim 10^9 \; \hbox{GeV}  \,, \quad M_\KK \sim 10^{12} \; \hbox{GeV}  \,, \quad M_s \sim 10^{14} \; \hbox{GeV} \quad \hbox{if} \quad |w_0| \sim M_p^3 \nn\\
   &&\qquad\tau_0 \sim 10^8 \,, \quad m_{3/2} \sim  M_\KK \sim 10^{10} \; \hbox{GeV}  \,, \quad M_s \sim 10^{12} \; \hbox{GeV} \quad \hbox{if} \quad |w_0| \sim M_p^3 \, \tau_0^{1/2} \,. \nn
\eea

Given a value for $\tau_0$, the size of soft supersymmetry-breaking terms, superpartner masses and trilinear couplings for any Standard Model like sector can also be estimated, under the assumption that their dominant source of supersymmetry breaking comes from the $T$ auxiliary field, although the result depends somewhat on the particular microscopic realization of the Standard Model and hidden sectors. For instance, suppose a Standard Model multiplet $\psi^i$ appears in the quantity $k(\psi, \ol\psi)$ of eq.~\pref{Kexpansion}. This would arise, for example, for states sequestered in local D3 or D7 branes and predicts soft supersymmetry-breaking masses in a manner similar, although slightly different dependence, to what is found for the large modulus in LVS (see for instance \cite{Conlon:2008wa}):
\be
m_\psi^2 = m_{3/2}^2-F^i F^{\ol j}\partial_i \partial_{\ol j}\ln Z_\psi \quad \hbox{which implies} \quad m_\psi\sim \frac{w_0}{\tau^2}\sim \frac{m_{3/2}}{\tau^{1/2}} \,.
\ee
Here $Z_\psi \sim \partial_{i \bar \jmath}K \sim - k_{i \bar \jmath}/\tau$ and the $F$-term for $T$ is given by
\be
F^\ssT=e^{K/2} K^{\ssT\ol \ssT}K_\ssT W\sim \frac{w_0}{\tau^{1/2}}+\cO(\tau^{-3/2}) \,.
\ee

For gaugino masses in this type of scenario it is instead the $F$-term of the dilaton that plays the key role. This is true (as in LVS) even though to leading order the dilaton $\cS$ does not break supersymmetry $F^S\propto D_SW=0$, since to next order in the $1/\tau$ expansion we have $F^\ssS\simeq e^{K/2}K^{\ssS\ol \ssS}K_\ssS w_0\sim w_0/\tau^{5/2}$. This gives gaugino masses of order
\be
M_\ssG=\frac{F^i\partial_i f}{{\mathrm Re}f}\sim \frac{w_0}{\tau^{5/2}}\sim \frac{m_{3/2}}{\tau} \,.
\ee
Similarly for the scalar trilinear soft-couplings in the potential ($A$ terms) which are $A\sim M_\ssG$. 

Unless further cancellations happen, this gives a split-SUSY spectrum in the sense that gauginos are lighter than scalar masses and both of these are lighter than the gravitino mass and similar to the mass of $\tau$. Since $\tau$ has to be heavier than $30$ TeV, this allows for gaugino masses of order the TeV scale for $\tau_0 \simeq 10^6$. Note however that we have only included the contribution  of the overall volume modulus $\tau$ and the dilaton $s$ to supersymmetry breaking, the contribution of the Standard Model cycle, if present, may dominate and wash out the sequestering making all soft terms of order the gravitino mass and leading to intermediate scale supersymmetry breaking. Therefore, similar to LVS but with different volume dependence, our scenario may lead either to sequestered split supersymmetry or intermediate scale supersymmetry breaking in both cases with intermediate scale gravitino mass.  A more detailed study of the structure of soft terms requires constructing concrete realizations of the Standard Model sector and is beyond the scope of this article.

\subsection{The Dine-Seiberg problem}

With the broad picture of the string embedding in place we can further comment on how this construction bears on the challenges posed by the Dine-Seiberg problem. These challenges come in two separate forms, each of which we discuss in turn.

\subsubsection*{Field expansions}

At its most basic the problem starts with the observation that expansion parameters are fields in string theory, and for any potential of the form 
\be
   V(\tau) = \sum_n \frac{V_n}{\tau^n} 
\ee
the condition $V'(\tau_0) = 0$ requires at least two terms of this series to have a similar size. How can this be consistent with the underlying expansion in $1/\tau$? Taken at face value this means stationary points of $V$ must occur outside the perturbative domain. 

There are several well-known ways to evade this argument. One such observes that perturbation theory need not break down if the first coefficient $V_0$ (or the first few) is for some reason unusually small. If it happens that $V_0/V_1 = \cO(\epsilon)$ for some $\epsilon \ll 1$ with all coefficients except $V_0$ having roughly the same size, then the stationary point $\tau_0$ of the series $V(\tau) = \tau^{-p}(V_0 + V_1 \tau^{-1} + \cdots )$ satisfies $V'(\tau_0) = 0 = -\tau_0^{-p - 1} [p\, V_0 + (p+1) V_1 \tau_0^{-1} + \cdots ]$ and so has a solution well-approximated by
\be \label{tauhierV}
   \frac{1}{\tau_0} \simeq - \frac{p \,V_0}{(p+1)V_1} = \cO(\epsilon) \,.
\ee
Precisely this argument is applied to the series in powers of $\alpha_g$ in \pref{coeffhier} to find stationary points $V'(\tau_0) = 0$ consistent with the condition $\alpha_g(\tau_0) \ll 1$. It is also the argument ultimately used in single-modulus KKLT models, for which $|w_0|$ must be assumed small in order to balance against terms like $\delta W \propto e^{-a T}$ within the context of an overall expansion in power of $1/\tau$.

The large-volume scenario (LVS) of string vacua works with a variation of this theme that requires the existence of at least two expansion parameters. In this case the potential arises as a multiple expansion of the form
\be
   V(\tau_1, \tau_2) = \sum_{mn} V_{mn} \, \varepsilon_1^m \, \varepsilon_2^n 
\ee
where $\varepsilon_i(\tau_1,\tau_2)$ are two small functions of the two independent moduli (and in practice $\varepsilon_1 = \tau_1^{-1/2}$ and $\varepsilon_2 = e^{-a \tau_2}$). In this case different terms in $\partial_{\tau_i} V = 0$ can balance against one another provided the $\varepsilon_i$ are similar in size, but now this can be consistent with both of them being small without having to assume special properties for the coefficients $V_{mn}$. It is noteworthy from this point of view that multiple moduli is the rule for the underlying Calabi-Yau spaces of interest, not the exception.

The stabilization mechanism used here adds a third way to evade this problem. In this case two expansion parameters are present but the stabilization occurs completely using a fixed order in $1/\tau$ since the potential has the form $V(\tau) \simeq U(\ln \tau)/\tau^4$, with $U$ a rational function of $\ln\tau$ arising due to its expansion in powers of $\alpha_g(\tau)$. The existence of a minimum for $V$ with respect to variations of $\tau$ then also requires terms at different orders in $\alpha_g$ to balance, and this is achieved consistent with $\alpha_g(\tau_0) \ll 1$ by using the assumption \pref{coeffhier} -- similar to the reasoning leading to \pref{tauhierV}. The new ingredient arises because the $\tau$ dependence embedded in $\alpha_g(\tau)$ itself comes as a series in $\alpha_g \ln\tau$, and solutions come with $\tau$ large enough that $\alpha_g \ln \tau$ need not be small even if $\alpha_g$ is. Nevertheless the magic of the renormalization group ensures that the solution can be computed reliably even if $\alpha_g \ln \tau \sim \cO(1)$ without assuming anything special about the coefficients $b_i$ in eq.~\pref{betafunctiontau}. 
 
\subsubsection*{Perturbation theory and the runaway}

A stronger claim is sometimes superimposed on the Dine-Seiberg problem within string theory. This claim (emphasised in particular by advocates of swampland hypotheses and de Sitter conjectures \cite{Obied:2018sgi, Palti:2019pca}) states that full control over perturbative expansions in $1/\tau$ and/or $1/s$ are only valid for $\tau$ and $s$ large enough to be in the runaway region, for which standard EFT methods break down because there is no static string vacua exist about which to perturb. 

Different versions of this argument involve two separate claims: 
\begin{itemize}
\item In one the objection is that any effective 4D description inevitably breaks down for large enough values of fields like $\tau$ because for large $\tau$ a tower of high-energy states descend into the low-energy theory and ruin its validity. This tower is not hypothetical in the case where $\tau$ is the volume modulus because the solution in the limit $\tau \to \infty$ is 10D flat space and the dangerous tower in question consists of the higher-dimensional Kaluza-Klein modes. 
\item In  the other the objection is that the state of the string-theory art only provides tools for studying static vacua and these do not allow comparison with the runaway region for which $\tau$ and $s$ are asymptotically large but not actually infinite. 
\end{itemize}
We argue here that neither of these objections need be that worrisome.

In one sense the first objection is simply true: the first nonzero Kaluza-Klein mass provides the UV scale above which any 4D EFT must break down, and it is also true that $M_\KK \to 0$ as $\tau \to \infty$. As argued in \cite{EFTBook} it is {\it never} a good approximation to include even a few KK states with nonzero masses in the 4D EFT while neglecting the rest because the underlying EFT expansion is in $M_{\rm low}/M_{\rm high}$ where $M_{\rm low}$ is the highest nonzero energy scale appearing in the low-energy theory and $M_{\rm high}$ is the lowest energy scale intrinsic to the high-energy theory. But Kaluza-Klein masses come in quantized towers, such as when $M_n = n/L$ for $n$ some integer and $L$ some extra-dimesional length. Keeping the $n=1$ state in the low-energy EFT while integrating out the $n=2$ state is only justified within an expansion in powers of $M_1/M_2 = \frac12$; never a parametrically small variable. 

The same problem does {\it not} arise if only $n=0$ states ({\it i.e.}~the moduli) are included in the 4D theory because in this case the low-energy mass need not be tied as rigidly to the KK scale $1/L$. Whether a 4D theory makes sense depends on the masses acquired by the moduli, and in the examples of interest here we have {\it e.g.}~$m_\tau/M_\KK \propto \tau^{-1} \sim (M_s L)^{-4}$, which can be parameterically small precisely when $\tau \gg 1$ because then $M_\KK \sim 1/L \ll M_s$. Although it is true that the UV cutoff of the 4D theory declines monotonically as $\tau \to \infty$, the 4D theory can have a nontrivial domain of validity for any large but finite $\tau$. One must of course check that this suffices to describe the physical process of interest.

The second objection ultimately puts a premium on static solutions when justifying using EFT methods. There is no evidence, however, that this is required elsewhere in the many areas of physics for which EFT methods apply. It is true that EFTs have additional conditions for validity when applied to time-dependent systems: most notably the motion must be adiabatic in the sense that $\dot \phi/\phi$ must be a low-energy scale ({\it i.e.}~be much smaller than UV scales like $M_\KK$) for any moving low-energy field $\phi$ (see \cite{EFTBook,Burgess:2020nec} for a more detailed discussion). But once these are satisfied the usual rules for EFTs apply and there is no necessity to expand around a strictly static vacuum solution (for examples where 4D effective evolution is compared to explicit higher-dimension evolution see \cite{Burgess:2016ygs}). In the end the only issue is whether the 4D EFT is being used within its domain of validity, including for applications to time-dependent problems. 

Many examples from other types of physics parallel the runaway situation encountered for fields like $\tau$ and $s$ in string theory (and extra-dimensional models more generally). One of these is the interaction energy of two nearby atoms regarded as functions of their centre-of-mass positions, $V(\bfr_1, \bfr_2)$. For spinless atoms the low-energy EFT variables can simply be the $\bfr_i$ if internal atomic size and structure define the UV scale.
 
It can happen that van der Waals forces can make atoms attract for large separations until more microscopic ({\it e.g.}~exchange) forces eventually intervene to convert this to strong repulsion. For specific atomic properties a stable static (molecular) solution can exist for a specific separation, $r := |\bfr_1 - \bfr_2| = a_0$, with no static solution possible for any other finite $r$. This does not forbid the use of EFT methods (such as the Born-Oppenheimer approximation) for values $r \neq a_0$, such as is done to map out the shape of the potential $V(\bfr_1, \bfr_2)$. Neither does it make expansion around the free theory (in powers of $1/r$ around $r = \infty$) useless. We see no reason why the large-$\tau$ and large-$s$ limits of string theory should not be as benign as is this everyday analog.

Simpler atomic systems also shed light on arguments that non-perturbative effects require a stable vacuum in order to be computable \cite{Sethi:2017phn} (which we mention despite our stabilization mechanism not requiring the use of non-perturbative effects).  For instance, suppose one or both of the underlying nuclei were to be unstable to $\alpha$ decay with a very long half-life (such as Uranium). Since $\alpha$-decay proceeds through the tunneling of He nuclei through a Coulomb barrier it can be regarded as a nonperturbative effect. It is hard to argue that one cannot compute -- even in principle -- the decay lifetime of the uranium atom except at the one place where the molecule is static. Notice that because the tunnelling rate is through an electromagnetic potential it depends in principle on the position of the external electrons and so at some small level also depends on the inter-atomic separation since the electrons adjust to the presence of the other atom in the molecule.

We conclude that even though string theory has no free parameters and weak couplings are related to runaway directions, the underlying issues of control are not unique to string theory. Experience with reliable calculations under similar conditions elsewhere in physics suggest that trustable perturbative calculations should also be reliable in string theory.

\section{Non-linear SUSY and Inflation}
\label{sec:NLSUSYInf}

de Sitter solutions enter practical considerations in one of two ways: as descriptions of our cosmological future or as descriptions of our distant inflationary cosmological past. Although the solution described above might conceivably describe the future universe, the necessity for inflation eventually to end means that it cannot in itself do so in the past. We next argue that minor extensions of the previous section's discussion can include both cases, along the way potentially explaining why each involves such different scales.

\subsection{The goldstino and the inflaton}

To incorporate inflation we require two new ingredients. We first require a large source of positive potential energy and because this necessarily breaks supersymmetry we require a sector that breaks supersymmetry more dramatically than does the dilaton multiplet\footnote{One might ask whether inflation can be obtained directly from motion driven by the $\tau$ potential \pref{VFBasicForm} itself. We have been unable to do so within a regime under EFT control, largely because the $\tau^{-4}$ behaviour of the potential is too steep.} $T$. The minimum number of degrees of freedom such a sector can introduce at low energy is the goldstone fermion, $G$, for supersymmetry breaking. We follow \cite{Komargodski:2009rz} and incorporate this fermion into the present discussion using a chiral field $X$ that satisfies a nilpotentcy constraint: 
\be \label{Xconstraint}
     X^2 = 0 \,. 
\ee
This constraint removes any scalar superpartners of $G$ and ensures that it nonlinearly realizes supersymmetry. The couplings of such fields to supergravity are explored in \cite{NilpotentSUGRA}. 

The second new ingredient is an inflaton field $\phi$ that can interpolate between a region where the large supersymmetry-breaking energy dominates and one where it does not. Inflation is then imagined to take place as the gravitational byproduct of the slow evolution of the field $\phi$ between these different regimes. With later applications to brane-antibrane inflation in mind we imagine $\phi$ also to arise within the sector for which supersymmetry is badly broken. A nonsupersymmetric scalar $\phi$ can also be represented by a chiral superfield $\Phi$ subject to a constraint \cite{Komargodski:2009rz, NilpotentSUGRA}, which in this case becomes:
\be \label{Phiconstraint}
 \ol \cD (X\ol \Phi)=0 \,.
\ee 
This states that $X \ol\Phi$ is left-chiral. If $\phi$ is also real then the left-chiral field it is equal to is $X\Phi$, in which case \pref{Phiconstraint} strengthens to the constraint $X(\Phi - \ol\Phi) = 0$. In either case the constraint removes the fermionic and auxiliary-field components of $\Phi$ in a way consistent with nonlinearly realized supersymmetry. 

To incorporate these fields into a supersymmetric framework with accidental approximate scale invariance we repeat the previous section's construction but now include these two new fields. For example, the K\"ahler potential built only from the minimal superfields $X$, $T$ and $\Phi$ is, as before,
\be\label{RelaxK2}
  e^{-K/3} = \tau - k + \frac{h}{\tau} + \cdots   \,,
\ee
where the ellipses denote higher orders in $1/\tau$, but now  
\be \label{kexpX}
  k =   \mfK(\Phi,\ol{\Phi},\ln \tau) +  (X + \ol{X}) \mfK_\ssX(\Phi,\ol{\Phi},\ln\tau) +  \ol{X} X \mfK_{\ssX\ol\ssX}(\Phi,\ol{\Phi},\ln\tau)  \,,
\ee
and similarly for $h$ and higher-order terms (although these are not needed in what follows). The most general superpotential similarly is
\be\label{RelaxW2}
   W \simeq w_0(\Phi) +  X w_{\ssX}(\Phi,\ol\Phi) \,,
\ee
where the unusual dependence of $W$ on $\ol\Phi$ is allowed because the constraint \pref{Phiconstraint} ensures that the result is chiral once multiplied by $X$.

The component lagrangian obtained from $K$ and $W$ is as given in \cite{NilpotentSUGRA}. The constraint \pref{Xconstraint} ensures there is no independent propagating scalar for the $X$ multiplet, but the kinetic terms for the scalar parts of the remaining fields $z^\ssI := \{ T, \Phi \}$ are given by the standard form
\bea \label{TkinLinf}
  - \frac{\cL_{\rm kin}}{\sqrt{-g}} &=& K_{\ssI \ol\ssJ} \, \partial_\mu z^\ssI \, \partial^\mu \ol z^\ssJ \nn \\
  &\simeq& \frac{3}{\cP^2} \left(1 + \frac{k'' - 2k'}{\cP} \right) \partial_\mu \cT\, \partial^\mu \ol \cT -  \left[ \frac{3}{\cP^2} \left(k_\phi - k_{\phi}' \right) \partial_\mu \phi\, \partial^\mu \ol \cT  + \hbox{h.c.}\right] \\
  && \qquad \qquad \qquad + \frac{3}{\cP} \left( k_{\phi\ol\phi} + \frac{k_\phi k_{\ol\phi}}{\cP} \right) \; \partial_\mu \phi \, \partial^\mu \ol\phi + \cdots \nn  \,,
\eea
where the ellipses denote terms involving higher-order coefficients like $h$ and primes denote derivatives with respect to $\ln \cP$.

The scalar potential resembles the standard supergravity form, although the absence of auxiliary fields in $\Phi$ also makes it slightly different. Writing $z^\ssA := \{ T, X\}$ and adopting the standard notation where $K^{\ol \ssA \ssB}$ is the inverse to the matrix $K_{\ssA \ol \ssB}$, the scalar potential works out to have the familiar supergravity form \cite{NilpotentSUGRA}
\be \label{VFdefinf}
  V = e^{K} \Bigl[ K^{\ol \ssA \ssB} \ol{D_\ssA W} D_\ssB W - 3|W|^2 \Bigr]   \,.
\ee
with the important proviso that the constrained field $\Phi$ is not included in the index sums over $A$ and $B$. $X = 0$ must be chosen (after differentiation) when tracking the dependence on scalar fields because \pref{Xconstraint} ensures the scalar part of $X$ is built from fermion bilinears. 

The scalar potential again comes as a series in inverse powers of $\tau$, whose leading terms turn out to be
\be \label{VABC}
   V =   \frac{A |w\ssX|^2}{\cP^2}-\frac{2{\rm Re}(B\, \ol{w_\ssX}  w_0)}{\cP^3}+\frac{C|w_0|^2}{\cP^4} \,,
\ee
where (as before) $\cP := \tau - k + \cdots$ and we keep in mind that each $T$ derivative of $k$ costs a power of $1/\cP$ because $k$ is a function of $\ln\cP$ rather than just $\ln\tau$ (see the observation in the footnote below eq.~\pref{alphavsmu}). The coefficients appearing in \pref{VABC} are given explicitly by
\be \label{ABCdefs}
  A \simeq \frac13 \, \mfK^{\ol\ssX \ssX} \,,  \quad
  \frac{B}{\cP} \simeq   \mfK^{\ol\ssX \ssX} \mfK_{\ssX \ol\ssT}   \quad \hbox{and} \quad
  \frac{C}{\cP^2} \simeq  - \frac{3(\mfK_{\ssT\ol\ssT}  - \mfK^{\ol\ssX \ssX} \mfK_{\ssT \ol\ssX} \mfK_{\ssX \ol\ssT})}{  1  + 2 \mfK^{\ssX\ol\ssX}  \mfK_\ssX \mfK_{\ol\ssX} } \,,
\ee
and we assume $A > 0$ so the leading $|w_\ssX|^2$ term is positive. Notice that \pref{VABC} reduces to \pref{VFBasicForm} in the limit $w_\ssX = k_\ssX = k_{\ssX \ol \ssT} = 0$. Notice also that $\mfK_{\ssT\ol\ssT}$ and $\mfK_{\ssX\ol\ssT}$ would both be $\cO(\alpha_g^2)$ if $k$ inherits its $T$ dependence through the $\cP$-dependence of a perturbative coupling $\alpha_g(\cP)$, as in \S\ref{ssec:taustab}, and this suggests that $B/A,C/A \sim \cO(\alpha_g^2)$. The inflationary implications of this potential are the subject of the remainder of this section.

\subsection{Inflaton potential}
\label{ssec:taustab2}

The potential \pref{VABC} is a function of the two scalars $\tau$ and $\phi$, and we turn now to exploring its properties. The first step is to make contact with previous sections, which can be done if there exists a configuration $\phi_0$ for which $w_\ssX(\phi_0) = 0$. When this exists it is very close to a minium of the potential, since it effectively turns off the leading $\cP^{-2}$ and $\cP^{-3}$ terms in $V$, leaving terms of order $|w_0|^2/\cP^4$ to dominate. $\tau$ can be minimized in this regime using the $\ln \cP$ dependence of $k$ very much along the lines discussed in \S\ref{sec:dS}. 

Within an inflationary perspective this minimum with $\phi$ near $\phi_0$ is the endpoint of any $\phi$ evolution and so represents our present-day vacuum. The inflationary regime instead occurs for $\phi$ far from $\phi_0$, where the potential is dominated by the large positive contribution coming from the $|w_\ssX|^2/\cP^2$ term in \pref{VABC}. The next sections compute how the solution for $\tau(\phi)$ obtained by minimizing $V(\tau,\phi)$ changes between these inflationary and post-inflationary regimes.

\subsubsection*{Minimization at late times}
\label{ssec:taustab3}

Suppose first that there exists a field configuration $\phi_0$ that satisfies $w_\ssX(\phi_0) = 0$, and consider the extremization problem near this point. This can be done very simply in the particular case where none of $\mfK^{\ol\ssX \ssX}$, $\mfK_\ssX$ and $\mfK_{\ssT\ol\ssX}$ depend on $\phi$ (though this assumption of $\phi$-independence can also be relaxed -- see \cite{YogaDE}). In this case $\phi$ enters the potential only through $w_\ssX$ and so minimizing $V$ with respect to $\phi$ amounts to doing so with respect to $w_\ssX$. 

Since $V$ is quadratic in $w_\ssX$ it is extremized by evaluating at the saddle point 
\be \label{wXsaddle}
   w_\ssX = 3 \mfK_{\ssX \ol\ssT}  \, w_0 \,,  
\ee
leading to 
\be \label{VFBasicFormzzmin22}
  V[\phi(\tau),\tau]  \simeq  -\frac{3|w_0|^2}{\cP^2 } \left[  \mfK^{\ol\ssX \ssX}   \mfK_{\ssX \ol\ssT}  \mfK_{\ssT \ol\ssX}  + \frac{\mfK_{\ssT\ol\ssT}  - \mfK^{\ol\ssX \ssX} \mfK_{\ssT \ol\ssX} \mfK_{\ssX \ol\ssT}}{  1  +2 \mfK^{\ssX\ol\ssX}  \mfK_\ssX \mfK_{\ol\ssX} }
 \right] =: \frac{U}{\cP^4}  \,.
\ee
Notice that this vanishes when $k$ is independent of $T$. As described in the footnote below eq.~\pref{alphavsmu}, we write the potential as a function of $\cP$ (rather than just $\tau$) because this plays an important role in our later discussion of the $\eta$ problem.  

The potential \pref{VFBasicFormzzmin22} now has precisely the same form considered in \S\ref{sec:dS}, though with $U = U(\ln \cP)$ given by a slightly different function of $k$ and its derivatives. Minimization with respect to $\tau$ can be carried over in whole cloth provided that $k$ depends on $\ln\cP$ in the way described in \S\ref{ssec:taustab}, such as if 
\be \label{mfKvsalpha}
   \mfK \simeq \mfK_0 + \mfK_1 \alpha_g + \frac{\mfK_2}{2}\, \alpha_g^2 + \cdots 
   \quad \hbox{and} \quad
   \frac{1}{\alpha_g} = b_0 - b_1 \ln \cP 
\ee
and so on (for constants $b_i$). The arguments of \S\ref{sec:dS} then go through as before, showing that a minimum is possible at $\cP = \cP_0$ with $\alpha_0 = \alpha_g(\cP_0) \sim \cO(\epsilon)$ and $\ln\cP_0 \sim 1/\epsilon$, with $\epsilon \ll 1$. The potential evaluated at this minimum is again order $V \sim \epsilon^5 |w_0|^2/\cP_0^4$ and so can be extremely small. The field $\phi$, introduced here as an inflaton, plays the role of a relaxation field by dynamically minimizing the $|w_\ssX|^2$ term (providing a simple rationale for the relaxation mechanism as applied in \cite{YogaDE} to present-day Dark Energy). 

\subsubsection*{Inflationary regime}

The other region of interest for the potential \pref{VABC} is when $\phi$ is far from $\phi_0$ and so $w_\ssX$ is not small. This is the regime likely appropriate for inflationary evolution since the potential is dominated by the large positive contribution proportional to $|w_\ssX|^2$. In general the fields $\tau$ and $\phi$ can evolve independently in this regime and it is not {\it a priori} necessary that either should sit at a local minimum of the potential \cite{Barreiro:2005ua}. However when seeking the conditions for slow-roll inflation it is instructive to first extremize the potential \pref{VABC} with respect to $\tau$ and ask how the result looks as a function of $\phi$. 

To this end, suppose that $\phi$ lies in a region for which $\delta := |w_\ssX / w_0| \lsim 0.01$ is small (a regime that actually arises for large $\phi$ for the brane-antibrane example described below because the brane tensions are parametrically smaller than the extra-dimensional Planck scale). In this case $V$ is a sum of terms of relative order $\delta^2$, $\delta/\cP$ and $1/\cP^2$, and so can be extremized for values $1/\cP \sim \delta$. For example, for real $B$, $w_\ssX$ and $w_0$ the extrema are $\tau = \tau_\pm(\ol\phi,\phi)$ where\footnote{For this we neglect for simplicity any dependence of $A$, $B$ and $C$ on $\ln\cP$ that is implicit through their dependence on $\alpha_g(\cP)$. This neglect is justified because $\alpha_g(\cP) = \alpha_{g0} /(1 - b_1 \, \alpha_{g0} \ln \cP)$ ensures that all $\ln \cP$-dependence is subleading in powers of $\alpha_{g0}$. This argument is consistent with keeping all orders in $\alpha_{g0} \ln\cP_0$ for the present-day stabilization of $\cP_0$ because $\cP$ turns out to be much smaller during inflation than at present.}
\be \label{taumininf}
    \frac{1}{\cP_\pm} \simeq \frac{ D_\pm w_\ssX}{w_0}  \sim \cO(\delta) \quad \hbox{with}\quad D_\pm:=   \frac{3B}{4C}  \pm \sqrt{\frac{9B^2 }{16C^2}-  \frac{A}{2C}}    \,.
\ee
This expression predicts several regimes:
\begin{itemize}
\item When $A/C < 0$ only one of these roots is positive, corresponding to a local maximum with $V_{\rm max} > 0$ when $A > 0$ (and a local minimum with $V_{\rm min} < 0$ when $A < 0$). 
\item If $A/C > 0$ then no stationary points exist at all for positive $\cP$ if $9\, B^2 < 8 AC$. For fixed $\phi$ the potential $V$ is then a monotonically decreasing positive function if $A > 0$. 
\item Both roots are real and positive (with $\cP_-$ being a local maximum and $\cP_+$ a local minimum) when $A/C$ and $B/C$ are both positive and $9B^2 > 8AC$. The potential evaluated at the minimum $\tau_+$ is also positive provided $B^2 < AC$ and negative otherwise.
\end{itemize}

When a local minimum exists and $A/C$ and $B/C$ are $\cO(1)$ and $\epsilon \sim \delta$ are similar in size (and small) then $\tau_+(\phi) \sim 1/\delta \gg 1$ is large but is much smaller than its value $\tau_0 \sim \exp(1/\epsilon)$ at the global minimum. If on the other hand $|A/C| \gg |B/C|$ -- as might be expected if $A \sim \cO(\alpha_g^0)$ and $B,C \sim \cO(\alpha_g^2)$ -- then stationary points only arise for positive $A$ if $C < 0$, corresponding to a local maximum with $V_{\rm max} > 0$. Notice that having $B/A, C/A \propto \alpha_g^2 \sim \cO(\epsilon^2)$ need not be inconsistent with the existence of two roots $\tau_\pm$ (and so having a local minimum) but only if the numerical coefficients in $C/A$ are adjusted to be $\cO(\epsilon^2)$, since then $B^2/C^2$ and $A/C$ are both $\cO(\epsilon^{-4})$. This can be done, for example, by arranging $\mfK_{\ssT\ol\ssT}$ to be numerically suppressed by $\epsilon^2$ so that $\mfK_{\ssT\ol\ssT} \sim \cO(\epsilon^2 \alpha_g^2)$ is similar in size to $(\mfK_{\ssX\ol\ssT})^2 \sim \cO(\alpha_g^4)$. Using this in \pref{taumininf} then shows that $\cP_\pm \sim \epsilon^2/\delta$ are only large if $\delta \ll \epsilon^2$.

In the special case where a local minimum exists we can compute an effective potential for $\phi$ defined by $ V_{\rm eff}(\phi) = V[\phi, \tau_+(\phi)]$, for which $\tau$ is assumed to remain at its local minimum as $\phi$ changes. This evaluates to
\be
\label{Veffnotau}
    V_{\rm eff}(\phi) \simeq \frac{(A-BD_+)}{2\cP_+^2}\, |w_\ssX(\phi)|^2 =\frac{(A-BD_+)D_+^2}{2|w_0|^2}\, |w_\ssX(\phi)|^4
\ee

\subsection{Inflationary evolution}
\label{ssec:infevo}
 
For inflation we are interested in how the fields $\phi$ and $\tau$ evolve and so must keep track of their kinetic terms given in \pref{TkinLinf}. Although in general $\tau$ and $\phi$ are mixed by the kinetic terms, this mixing arises at subleading order in $1/\tau$. The leading form for the target-space metric is $K_{\ssI \ol\ssJ} \, \exd z^\ssI \bar z^\ssJ \sim (\exd \tau/\tau)^2 + k_{\phi\ol\phi} |\exd \phi |^2/\tau$ and so for $k_{\phi\ol\phi} \sim 1$ the canonical fields are $\exd\chi \sim M_p \, \exd \tau/\tau$ and $\exd \varphi \sim \exd\phi/\sqrt{\bar\tau}$ near a semiclassical background $\tau = \bar\tau$. 

In general, motion along the $\tau$ direction is not a slow roll if inflation is dominated by the term $V \sim |w_\ssX|^2/\tau^2$ since the slow-roll parameter in the $\tau$ direction is then
\be
   \varepsilon_{(\tau)} \sim \left( \frac{M_p V_\chi}{V} \right)^2 \sim \left( \frac{ \tau V_\tau}{V} \right)^2 \sim \cO(1) \,.
\ee
The $\tau$ `mass' can be similarly estimated when $V \sim |w_\ssX|^2/\tau^2$ dominates, giving
\be
  m_\tau^2 = \left(\frac{\partial^2 V}{\partial \chi^2}\right)_{\tau_+} \sim \frac{\tau^2}{M_p^2} \frac{\partial^2 V}{\partial\tau^2}  \sim \frac{V}{M_p^2} \sim H_\ssI^2 \,,
\ee
where $H_\ssI$ is the inflationary Hubble scale, during an epoch when $V$ dominates the gravitating energy density. This is also insufficiently small to justify a slow roll. 

This makes it important to choose parameters so that $V$ has a local minimum in the $\tau$ direction that stabilizes $\tau = \tau(\phi)$, in which case we can see whether motion in the $\phi$ direction can be sufficiently slow. Even if this can be done it must be asked whether it is a good approximation to have $\tau$ remain trapped at its local minimum as $\phi$ evolves. This depends on whether the volume modulus is heavy enough to integrate out the transverse $\tau$ field to obtain an effectively single-field description. A brief estimate arguing that it is heavy enough is given in Appendix \S\ref{AppTrap}.

Evaluated at the local minimum for $\tau = \tau_+(\phi)$ described above we can evaluate the slow-roll parameters for evolving in the direction of the canonically normalized field $\varphi$, and this is particularly simple when $\phi$ dominantly enters through $w_\ssX$, since
\be \label{Vderivephi}
   \frac{\partial V}{\partial \varphi} \simeq \frac{\partial}{\partial\varphi} \left(\frac{ A |w_\ssX|^2}{\tau_+^2} - \frac{2\hbox{Re} \, (B \ol{w_\ssX} w_0)}{M_p^2\tau_+^3} + \frac{C|w_0|^2}{M_p^4\tau_+^4} \right)
   \simeq \left( \frac{A w_\ssX}{\tau_+^2} - \frac{B w_0}{M_p^2\tau_+^3} \right) \frac{\partial \ol w_\ssX}{\partial \varphi} + \hbox{h.c.} \,,
\ee
where factors of $M_p$ are re-instated for later convenience. Evaluating at $\tau=\tau_+(\phi)$ ensures each term in $V$ and $\partial V/\partial \varphi$ has a comparable size and so ensures $w_\ssX \sim (B w_0)/(AM_p^2\tau_+)$ and so we may estimate the right-hand side of \pref{Vderivephi} as being of order $[B w_0/(M_p^2 \tau_+^3)] (\partial w_\ssX/\partial \varphi)$.  For comparison, with these same estimates the potential energy itself is $V \sim C |w_0|^2/(M_p^4 \tau_+^4)$ and the inflationary Hubble scale is $H_\ssI \sim \sqrt{V}/M_p \sim \sqrt{C}\, |w_0|/(M_p^3 \tau_+^2)$. 

The first slow-roll parameter for motion in the $\phi$-direction then becomes
\be\label{firstepsilon}
   \varepsilon = \frac12 \left( \frac{M_p \, \partial V/\partial \varphi}{V} \right)^2 \sim \left( \frac{BM_p^3 \tau_+}{C |w_0|} \, \frac{\partial w_\ssX}{\partial \varphi} \right)^2 \sim \tau_+^3 \left( \frac{BM_p^3}{C |w_0|} \, \frac{\partial w_\ssX}{\partial \phi} \right)^2 \,,
\ee
which shows that slow roll requires $\partial w_\ssX/\partial \phi$ to be much smaller than order $C|w_0|/ (BM_p^3 \tau_+^{3/2})$. The second slow-roll parameter is similarly estimated to be
\bea\label{firsteta}
 \eta =  \frac{M_p^2}{V} \, \frac{\partial^2 V}{\partial \varphi^2}  &\sim& \frac{1}{H_\ssI^2} \left[ \frac{B |w_0|}{M_p^2 \tau_+^3} \left(\frac{\partial^2 w_\ssX}{\partial \varphi^2} \right) + \frac{A}{\tau_+^2} \left( \frac{\partial w_\ssX}{\partial \varphi} \right)^2  \right] \nn\\
 &\sim& \tau_+\left[ \frac{BM_p^4 \tau_+}{C |w_0|} \left(\frac{\partial^2 w_\ssX}{\partial \phi^2} \right) + \frac{AM_p^6 \tau_+^2}{C|w_0|^2} \,  \left( \frac{\partial w_\ssX}{\partial \phi} \right)^2  \right]\,.
\eea
The property $B^2 \sim AC$ -- required for the existence of a minimum $\tau_+(\phi)$ -- ensures the second term of this expression is small when $\varepsilon$ is small, so requiring small $\eta$ implies $\partial^2 w_\ssX/\partial \phi^2$ is much smaller than order $C |w_0|/(B M_p^4\tau_+^2)$. Using $w_0 \sim M_p^3$ and $(B/A)^2 \sim C/A \sim \epsilon^4 M_p^2$ and $\tau_+ \sim \epsilon^2/\delta$ (as found above when $B,C \propto \alpha_g^2$) then shows that slow roll requires the derivatives of $w_\ssX$ to satisfy
\be \label{infcondwx}
  \left| \frac{\partial w_\ssX}{\partial \phi} \right| \ll  \frac{C|w_0|}{BM_p^3 \tau_+^{3/2}} \sim  \frac{\epsilon^2 M_p}{\tau_+^{3/2}} \quad \hbox{and} \quad
  \left| \frac{\partial^2 w_\ssX}{\partial \phi^2} \right| \ll \frac{C|w_0|}{BM_p^4\tau_+^2} \sim \frac{\epsilon^2}{\tau_+^{2}}  \,.
\ee

To go further we must specify in more detail how the functions $k$ and $W$ depend on $\phi$. 
We next identify a promising choice for these functions by re-examining the brane-antibrane inflationary scenario  to see how it is affected by the new modulus stabilization mechanism.

\subsection{Warped D3-$\ol{\hbox{D3}}$ inflation revisited}
\label{ssec:WBABinf}

Brane-antibrane inflation was the first attempt to derive inflation from a string theory construction within a framework in which the inflaton potential could be explicitly calculated \cite{Burgess:2001fx,Dvali:2001fw} (see \cite{Kallosh:2018zsi} for a recent discussion). The inflaton field is the separation between a brane and an antibrane and for large separations the corresponding potential is the sum of two terms: the brane tension and the brane-antibrane interaction generated by their couplings to the various bulk fields. At large distances the inter-brane force takes a `Coulomb' form\footnote{We take the Coulomb energy for sources separated by a distance $y$ in $d$ transverse dimensions to be an interaction energy that falls like $y^{-p}$ with $p = d-2$  (with $p=4$ -- corresponding to $d=6$ -- being the case relevant to space-filling 3-branes in 10D string vacua).  } (and so weakens at large separations) but the challenge was to find how to separate the branes sufficiently within a finite-sized manifold and to compute how brane motion changes the energetics of modulus stabilization.
 
Let us briefly recall the main ideas. Fluxes in IIB compactifications back-react on the metric in such a way that the resulting compactification is a conformal Calabi-Yau threefold with metric of the form \cite{Burgess:2001fx,Dvali:2001fw,KKLMMT}
\be
   \exd s^2 = \left(1+\frac{e^{4\cA}}{\cV^{2/3}}\right)^{-1/2} \exd s_4^2+\left(1+\frac{e^{4\cA}}{\cV^{2/3}}\right)^{1/2} \exd s^2_{\ssC\ssY}
\ee
with $\cA(y)$ a calculable function of position within the extra dimensions. The warp factor $\cW := \left(1+{e^{4\cA}}{\cV^{-2/3}}\right)^{-1/2}$ plays a significant role in highly warped regions, defined by the condition $e^{4\cA}\gg \cV^{2/3} \gg 1$. 
For future use we note in passing that in order to have the warped string scale be larger than the Kaluza-Klein scale (as required to have a reliable low-energy effective field theory) the warping must be constrained to satisfy  \cite{WarpingBound}
\be \label{warpbound}
    e^{\cA} \lsim \cV^{2/3} \,.
\ee

A space-filling D3 brane sits at a particular point in the extra dimensions and experiences no position-dependent forces due to supersymmetric BPS cancellation of bulk forces, and so are free to move within the Calabi-Yau space. Anti-D3 branes by contrast energetically prefer to minimize the warp factor and so move to the tip of any warped throat for which $e^{4\cA}$ takes its largest value, which turns out to be
\be
e^{4\cA_{\rm tip}}:=e^{4\rho}=e^{{8\pi K}/({3g_sM})}
\ee
where $K$ and $M$ are the integer flux quantum numbers that fix the relevant complex structure moduli. Depending on the values of $K$ and $M$ the warp factor can be significant and so can naturally lead to a source of hierarchies within the theory.

Keeping in mind that $\tilde g_{\mu\nu} = \cW g_{\mu\nu}$ implies $\sqrt{-\tilde g_4} = \cW^2 \sqrt{-g_4}$, the tension of an anti-D3 brane localized at the tip of a strongly warped throat contributes the following positive contribution to the low-energy 4D scalar potential,
\be \label{TensionForm}
   2T_3 \cW^2 = \frac{ 2T_3}{1+ (e^{4\rho}/\cV^{2/3})}\simeq 2M_s^4 e^{-4\rho} \cV^{2/3} \sim \frac{e^{-4\rho}M_p^4}{\cV^{4/3}}  
\ee
which uses $e^{4\rho}\gg \cV^{2/3}$ for strongly warped regions as well as the value of the brane tension $T_3 \propto M_s^4$ and the relation \pref{MsMKK} between the string scale and the 4D Planck scale: 
\be \label{T3PlanckUnits}
   T_3=\frac{1}{8\pi^3 g_s\alpha'^2} = \frac{(2\pi)^{11}g_s^3M_p^4}{4\cV^2} \,.
\ee
The exponential dependence of the warp factor appearing within this brane tension is used in \cite{KKLT} (with a volume dependence later corrected by \cite{KKLMMT}) to uplift the AdS minimum found in previous modulus-stabilization mechanisms in order to obtain a dS solution rather than AdS or the more generic runaway. 

Combining this brane tension term with the Coulomb interaction between a mobile D3 brane and the anti-D3 brane sitting in a warped environment gives the candidate brane-antibrane inflation potential \cite{Burgess:2001fx,Dvali:2001fw,KKLMMT} (again in the Einstein frame):
\be\label{coulomb}
V= 2T_3 (e^{-4\rho}\cV^{2/3})\left(1-\frac{27}{64\pi^2}\frac{2T_3(e^{-4\rho}\cV^{2/3})}{|\varphi|^4}\right) =: \Omega\left(1-\frac{\mfb\, \Omega}{|\varphi|^4}\right)
\ee
where $\varphi$ is the canonically normalised field determining the brane separation $y$: $\varphi=\sqrt{T_3}\, y$ and the last equality evaluates $T_3$ in Planck units using \pref{T3PlanckUnits}, and so
\be\label{omega}
\Omega=\frac{\mfc \, e^{-4\rho}M_p^4}{\cV^{4/3}}, \quad  \mfc = \frac{(2\pi)^{11}g_s^3}2  \quad 
\hbox{and} \quad \mfb=\frac{27}{64\pi^2} \,.
\ee

\subsubsection*{Naive inflationary analysis}

Putting aside for the moment how $\cV$ evolves given this potential, consider first the naive single-field inflationary picture that emerges for $\varphi$ evolution at fixed $\cV$. The slow-roll parameters for this motion in the regime $\mfb \, \Omega \ll |\varphi|^4$ are
\be
  \varepsilon =  \frac{M_p^2}{2}\left(\frac{V_\varphi}{V}\right)^2\simeq 8\mfb^2\left(\frac{\Omega M_p}{|\varphi|^5}\right)^2
  \quad \hbox{and} \quad
  \eta =  \frac{M_p^2V_{\varphi\varphi}}{V} \simeq - \frac{20\mfb\,\Omega M_p^2}{|\varphi|^6} \,.
\ee
Although these can be made arbitrarily small by making $|\varphi|$ sufficiently large, as noted in \cite{Burgess:2001fx} inflation does not work (without the warp factors) because it would require the brane separation to be larger than the typical linear extent of the extra dimensions. But if the warp factors buried in $\Omega$ are small enough the slow roll conditions $\varepsilon\ll 1$ and $\eta\ll 1$ can be satisfied. Notice that these also imply that the ratio 
\be
  - \frac{\varepsilon}{\eta} \simeq  \frac{2\mfb \, \Omega}{5 |\varphi|^4} \ll 1\,,
\ee
is deep into the regime where quantum effects are dominated by stochastic methods \cite{Vennin:2015hra}. 

In terms of these the number of inflationary $e$-foldings between horizon exit and inflation's end is
\be
N_e=\frac{1}{M_p}\int_{\varphi_{end}}^{\varphi_*}\frac{d\varphi}{\sqrt{2\varepsilon}} \simeq \frac{\varphi_*^6}{24\mfb\, \Omega M_p^2}
\ee
and the amplitude of primordial scalar density perturbations becomes
\be
\delta_\ssH=\frac{1}{\pi\sqrt{75} } \left( \frac{V^{3/2}}{M_p^3\,V_\varphi} \right)_{\varphi_*} \simeq  \frac{\varphi_*^5}{4\pi\mfb\sqrt{75}M_p^3\sqrt{\Omega}} \,.
\ee
In these expressions $\varphi_*$ is the inflaton position at horizon exit, relative to which its value at inflation's end is neglected: $\varphi_* \gg \varphi_{\rm end}$. The slow-roll parameters evaluated at horizon exit then become 
\be \label{HEslowroll}
\eta_*=-\frac{5}{6N_e}, \qquad \varepsilon_* = \frac{20\pi \delta_\ssH}{9\sqrt{2} \, N_e^{5/2}}  \simeq \frac{16\pi }{5} \sqrt{\frac35} \, \delta_\ssH\, |\eta_*|^{5/2} \,.
\ee

The spectral index $n_s$ and tensor-to-scalar ratio $r$ are given by the usual expressions
\be\label{obsslowroll}
n_s=1+2\eta_*-6\varepsilon_* \simeq 1+ 2\eta_* \quad \hbox{and} \quad r= 16\varepsilon_*
\ee
in which $\varepsilon_*  \ll |\eta_*|$ is used in $n_s$, showing that the measured value for $n_s$ fixes $\eta_* \simeq \frac12(n_s - 1) \simeq - 0.015$ and so $N_e \simeq 56$. Combining \pref{HEslowroll} and \pref{obsslowroll} and using the measured amplitude $\delta_\ssH=1.9\times 10^{-5}$ then gives the following prediction for the scalar-to-tensor ratio
\be
   r = 16 \varepsilon_* \simeq \frac{64 \pi}{5} \sqrt{\frac{3}{10}}\, \delta_\ssH |n_s - 1|^{5/2} 
   \simeq 2 \times 10^{-8} \,,
\ee
which is too small to be observable in the foreseeable future.

Although at face value the warp factors (buried in $\Omega$) allow a potential as flat as desired, this assumes that the physics that stabilizes the overall volume modulus appearing in $\Omega$ has been fixed in a way that does not significantly alter the potential for $\phi$. However, the same shallowness that makes \pref{coulomb} attractive for inflation also makes it fragile to changes associated with modulus stabilization, as can be most clearly seen by embedding the brane-antibrane dynamics into a full 4D supergravity EFT that allows a consistent description of both inflaton and modulus-stabilization. As argued in \cite{KKLMMT} this exercise opens up a new problem (the $\eta$ problem) that generically ruins the shallowness of the potential \pref{coulomb}. We repeat this exercise here to show why the RG stabilization mechanism avoids this problem.

\subsubsection{The Nilpotent Superfield and anti-D3 Branes}
\label{sssec:nilpotentD3}

An interesting feature of the nilpotent superfield formalism of \S\ref{ssec:infevo} is that it captures very efficiently the physics of anti-D3 branes at the tip of a Calabi-Yau throat as described above in this section. We now explore this connection and determine the choices that it implies for quantities like $\mfK$, $w_0$ and $w_\ssX$. 

\subsubsection*{Antibrane tension from the nilpotent superfield}

First we recall how the $\cP$-dependence of the leading part of the potential built using a nilpotent superfield reproduces the volume dependence of the anti-D3 brane tension at the tip of the warped throat. For this recall that when $W = w_0 + w_\ssX X$ then the leading term in the scalar potential \pref{VABC} is
\be
V  \simeq \frac{\mfK^{\ol\ssX \ssX}|w_\ssX|^2}{3\cP^2}
\ee
which for $w_\ssX\propto e^{-2\rho}$ reproduces the KKLT expression \pref{TensionForm} once the volume modulus is identified in the usual way: $\cP = \cV^{2/3}$. 

It is noteworthy that this agreement between the volume-dependence of the nilpotent potential and the brane tension works only when one uses the warping-corrected volume-dependence given in \cite{KKLMMT} rather than the original expression of \cite{KKLT} that does not include warping. Only in the warped case is supersymmetry breaking sufficiently sequestered to be captured using only the single goldstino field $X$. From the string point of view the fact the $X$ has no independent scalar degree of freedom corresponds to the fact that the isolated anti-D3 brane has no position modulus because it is energetically stuck at the tip of the warped throat.

There is also additional evidence that warping can be captured by the superpotential in this way, since it can also be derived within the 4D effective supergravity as the expectation value of the throat's complex structure modulus $Y$, which schematically contributes to the superpotential in the form $W(Y)=Y(n_1\log Y+n_2)+YX$ with $n_i$ being integer flux quantum numbers. Eliminating $Y$ from this superpotential in a supersymmetric way gives rise to the required warp factor multiplying $X$ in $W$ \cite{Giddings:2001yu, Dudas:2019pls}. 

\subsubsection*{Inter-brane dynamics}

To obtain a supergravity representation for the dynamics of brane-antibrane motion we require a supermultiplet for the inter-brane separation field $\phi$. We do so here by representing the inter-brane separation using the constrained inflaton field $\Phi$ of \S\ref{ssec:infevo}. 

To describe the kinetic energy of these fields we choose the function $\mfK$ defined in \pref{kexpX} to have the form
\be \label{kexpX2}
   \mfK(\phi,\ol{\phi},\ln \cP) \simeq  \gamma(\ol \phi, \phi)+ \hat\mfK(\ln \cP) \,,
\ee
where $\phi^i$ is proportional to a complex coordinate describing the 3-brane position within the extra dimensions. The quantity $\hat\mfK$ is the $\phi$-independent function of $\ln\cP$ described in \pref{mfKvsalpha} whose presence stabilizes $\cP$ at exponentially large values in the present-day vacuum. In terms of this the kinetic term for changes to $\phi$ become
\be\label{phikinterm}
   - \frac{\cL_{\rm kin}}{\sqrt{-g}} = K_{i \bar \jmath} \, \partial_\mu \bar\phi^j \, \partial^\mu \phi^i 
   \simeq \frac{3\gamma_{i \bar \jmath}}{\cP} \, \, \partial_\mu \bar\phi^j \, \partial^\mu \phi^i 
\ee
showing that $\gamma_{i\bar\jmath}$ is naturally proportional to the extra-dimensional metric $g_{i \bar\jmath}$ and so $\gamma(\ol \phi, \phi)$ is proportional to this metric's K\"ahler potential. Coordinates can be chosen without loss of generality so that $\gamma \simeq \ol \phi \, \phi$ near $\phi = 0$. 
 
To capture the antibrane tension and the separation-dependent Coulomb interaction we use the following superpotential\footnote{See \cite{Aparicio:2015psl} for a first  supersymmetric discussion of the Coulomb potential.} 
\be \label{doublew}
W=w_0+Xw_\ssX(\Phi, \ol\Phi) \quad \hbox{with} \quad  w_\ssX(\Phi, \ol\Phi) = \mft - \frac{\mfg}{|\Phi|^4} + \cdots
\ee
where the ellipses denote terms suppressed by even higher powers of $|\Phi|^{-1}$.

With this choice the leading term in the scalar potential \pref{VABC} then is
\be \label{VcoulW}
V=\frac{\mfK^{\ol\ssX\ssX} |w_\ssX|^2}{3\cP^2} 
 = \frac{\mfK^{\ol\ssX\ssX}}{3\cP^2} \left[ |\mft|^2 -\frac{2\hbox{Re} (\ol\mft \mfg)}{|\phi|^4}+ 
  \cdots \right] \,,
\ee
which is to be compared to \pref{coulomb} and \pref{omega}. This comparison is most easily done using the canonically normalized field -- see \pref{phikinterm} -- $\varphi \propto \cP^{-1/2} \phi \propto \cV^{-1/3} \phi$, for which $|\phi|^4 \propto |\varphi|^4 \cP^2 \propto |\varphi|^4 \cV^{4/3}$\footnote{Note that it is not trivial that the volume-dependence in the scalar potential comes out  as required given the fact that the volume cannot appear in the superpotential.}. Once expressed in terms of $\varphi$ both terms of \pref{VcoulW} have the same dependence on $\cP$ (or $\cV)$ as in \pref{coulomb} and \pref{omega}, and so it becomes possible to read off the warping dependence of the coefficients $\mft$ and $\mfg$ in $w_\ssX$, leading to $\mfK^{\ol\ssX\ssX} |\mft|^2 \propto \mfc e^{-4\rho}$ and $\mfK^{\ol\ssX\ssX}2\hbox{Re}\, \ol \mft  \mfg \propto \mfb \mfc^2 e^{-8\rho}$. The freedom to rescale $X$ allows the warping dependence to be moved around somewhat, but if this is used to ensure $\mfK^{\ol\ssX\ssX}$ is warping free then (for real $\mft$ and $\mfg$) it implies
\be
 \mft \propto e^{-2\rho} \quad \hbox{and} \quad
 \mfg \propto e^{-6\rho} \,.
\ee

As mentioned earlier, it is a good thing that both $\mft$ and $\mfg$ are suppressed by warping because we have seen -- {\it c.f.}~eq.~\pref{infcondwx} -- that inflationary slow roll can be ensured by making the $\varphi$ derivatives of $w_\ssX$ sufficiently small. The subtlety in this argument is that $V$ is really a function of two fields, $\phi$ and $\tau$, and one must check that $\tau$ evolution does not ruin the desired inflationary behaviour. 

\begin{figure}[t]
\begin{center}
\includegraphics[width=100mm,height=60mm]{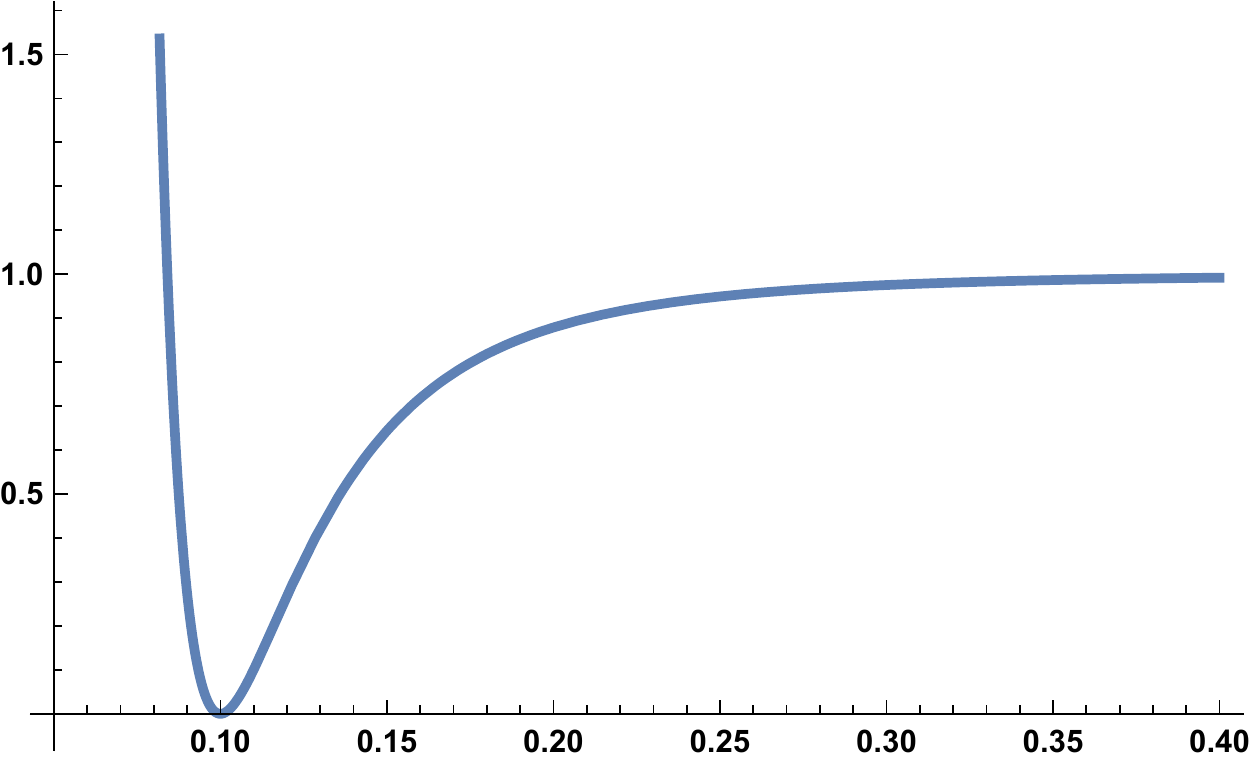} 
\caption{The Coulomb potential from the superpotential $w_\ssX(\phi)$. Since $V\propto |w_\ssX|^2$ this potential reproduces the Coulomb interaction for large values of the brane separation $\phi$ but at smaller distances the interaction potential has a minimum and becomes repulsive. At large field values, and due to the large amount of warping the potential is flat enough to give rise to inflation in a natural way. The minimum of the potential lies outside the domain of validity of the EFT and only for large values of $\phi$ is this potential under control.} \label{Fig:Coulomb} 
\end{center}
\end{figure}

The above discussion ignores the $|\mfg|^2/|\phi|^8$ term in $V$ because this must compete with higher order terms in $w_\ssX$ that are already dropped in \pref{doublew}.  It is noteworthy that some things can nonetheless be said for smaller $|\phi|$ even if the detailed form of $w_\ssX$ is not known in this limit. The main observation is that the supergravity structure of the leading $1/\cP^2$ term of $V$ comes proportional to $|w_\ssX|^2$ and so cannot decrease without bound as $|\phi|$ decreases. The lowest it can get is zero, which would be obtained if there exists a $\phi = \phi_0$ for which $w_\ssX(\phi_0) = 0$. If such a field exists within the domain of validity of the 4D EFT then this point is a local minimum of the potential; unlike the standard Coulomb interaction $V$ would reach a minimum value beyond which the interaction becomes repulsive rather than attractive --- see figure \pref{Fig:Coulomb}. Although in the simplest scenario this would usually be expected only to occur at scales of order the warped string scale or less -- and so be beyond the domain of 4D methods --  it remains to be seen whether more complicated examples exist for which $\phi_0$ can lie within the domain of 4D methods.

\subsubsection{Inflation, modulus stabilization and the $\eta$ problem}

The problem is that the $\tau$-dependence of $V$ {\it does} tend to ruin inflation, at least when $\tau$ is stabilized using the superpotential. To see why we now repeat the argument given in \cite{KKLMMT} that shows why stabilizing the volume modulus often introduces new issues. The problem is that the mechanism used in \cite{KKLMMT} to fix the volume modulus also induces a mass term for the would-be inflaton field $\varphi$ that makes the slow-roll parameter $\eta$ of order one. This is a special case of a more general problem for supergravity-based inflationary models, called the $\eta$ problem.

\subsubsection*{The $\eta$ problem}

The problem arises because the K\"ahler potential very generally depends on both $\tau$ and $\phi$: $K = -3 \ln[\tau - k(\phi, \bar \phi)+ \cdots]$, where $k(\phi, \bar\phi) \simeq \bar\phi\, \phi + \cdots$ is responsible for the kinetic term for $\phi$. So once $\tau$ is fixed by adding a holomorphic non-perturbative superpotential $W_{np}(T)$, the dependence of $K$ on $\phi$ introduces a potential energy that generates a mass for $\phi$ because of the potential's overall dependence on $e^K$:
\be\label{etaprobV}
   V = e^K \, \widehat V_0 \simeq \frac{\widehat V_0}{[\tau -\bar\phi\, \phi  + \cdots]^{3}} \simeq \frac{\widehat V_0}{\tau^3} \left[ 1 + \frac{3\bar\phi\,\phi}{\tau} + \cdots \right]  \simeq \frac{\widehat V_0}{\tau^3} \Bigl[ 1 + \bar\varphi\,\varphi + \cdots \Bigr] \,.
\ee
It is the superpotential terms within $\widehat V_0$ that contain the small warp factors that allow $\widehat V_0$ to depend so weakly on $\phi$ that inflation can be possible. The value of $\widehat V_0$ also fixes the value of the Hubble scale whenever the universal energy density is dominated by $V$, since then $H_\ssI^2 \simeq V/M_p^2 \simeq \widehat V_0/(\tau^3 M_p^2)$. But when this is so eq.~\pref{etaprobV} shows (once the $M_p$ factors are reinstated) that $\phi$ inevitably has a mass contribution that is of order $m^2_\phi \sim \widehat V_0/(\tau^3 M_p^2) \sim H^2_\ssI$ which therefore contributes a factor of order unity to the second slow-roll parameter $\eta = M_p^2 \,V_{\varphi\varphi}/V \simeq m_\phi^2/H_\ssI^2$. Slow roll is only achieved in the standard construction by including a large (unwarped) $\bar\phi\,\phi$ contribution into $\widehat V_0$ and tuning this to cancel against the term coming from $e^K$. Even though inflation is achievable in this way, it needs a very particular fine tuning and the Coulomb potential is essentially replaced by a tuned inflection-point inflation \cite{Baumann:2006th, Baumann:2007ah}. 

\begin{figure}[t]
\begin{center}
\includegraphics[width=100mm,height=60mm]{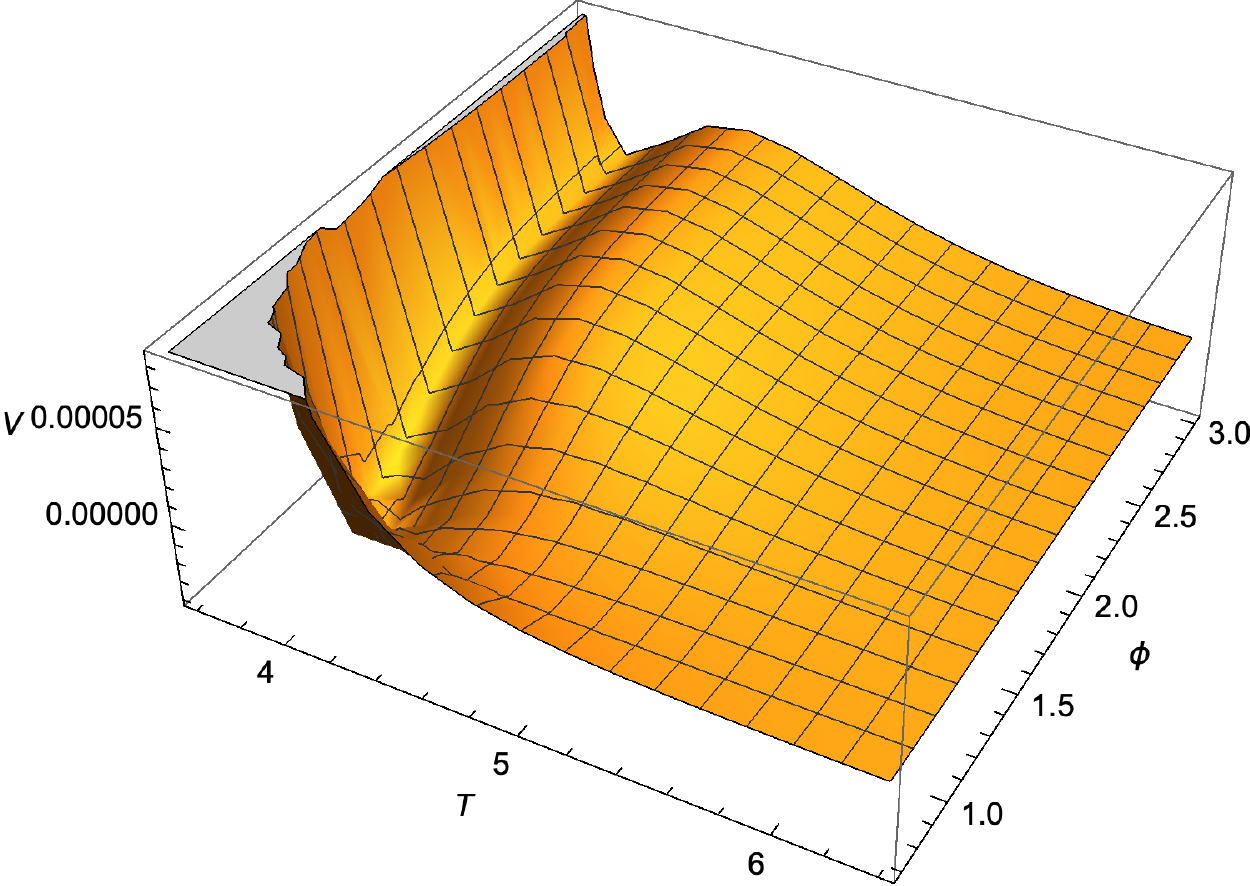} 
\caption{A plot showing the inflationary region in which the potential is temporarily fixed at a minimum in the modulus $\tau$ direction while the inflaton $\varphi$ slowly rolls. At the end of inflation the two fields run to their minimum at $\tau_0,\varphi_0$ with $\tau_0\gg \tau_{\rm inf}$.} \label{Fig:Inflation} 
\end{center}
\end{figure}

The problem is quite generic because the K\"ahler potential very generally depends only on $\cP = \tau - k + \cdots$ but because the superpotential must be a holomorphic function it cannot depend on $\cP$ and must only depend on $T$ and $\phi$ separately. But -- as already pointed out in \cite{KKLMMT} -- this also shows that it is potentially evaded if the modulus-stabilization mechanism can arise from corrections to $K$ rather than to $W$, provided these directly stabilize $\cP$ rather than just $\tau$. The modulus-stabilization mechanism presented here evades the $\eta$-problem in precisely this way: the stabilizing potential naturally arises as a function of $\cP$ directly. Inflation is therefore driven by an expression like \pref{etaprobV} regarded as a function of $\phi$ for fixed $\cP$ rather than for fixed $\tau$ (see Fig.~\ref{Fig:Inflation}).

\subsubsection*{Inflating with K\"ahler stabilization}

We now return to the combined $\tau$ and $\phi$ dynamics when $W$ is independent of $T$ and modulus stabilization instead arises through RG stabilization, with the potential generated using \pref{kexpX2}. There are two ways to proceed. The first simply uses the analysis given in \S\ref{ssec:infevo}, which explicitly follows the potential as a function of $\tau$ and $\phi$, specializing it to the superpotential \pref{doublew} and K\"ahler function \pref{kexpX2}. This leads to expressions \pref{firstepsilon} and \pref{firsteta} for the slow-roll parameters, and to the condition \pref{infcondwx} for inflation, which notably translates into the following conditions that really {\it can} now be satisfied by choosing sufficient warping
\be \label{infcondwx2}
  \frac{4\mfg}{|\phi|^5} \ll  \frac{C|w_0|}{BM_p^3 \tau_+^{3/2}} \sim  \frac{\epsilon^2 M_p}{\tau_+^{3/2}} \quad \hbox{and} \quad
   \frac{20\mfg}{|\phi|^6}  \ll \frac{C|w_0|}{BM_p^4\tau_+^2} \sim \frac{\epsilon^2}{\tau_+^{2}}  \,.
\ee

The second approach uses the effective potential $V_{\rm eff}(\phi) = V[\phi,\tau_+(\phi)]$ given in \pref{Veffnotau} obtained by evaluating $\tau = \tau_+(\phi)$ at its $\phi$-dependent minimum. Single-field methods are then directly applicable. Writing \pref{Veffnotau} as
\be\label{Veffnotau2}
    V_{\rm eff}(\phi) \simeq \frac{\mfD |w_\ssX(\phi)|^4}{|w_0|^2} \,,
\ee
we can directly differentiate to evaluate the slow-roll parameters. For these purposes we can neglect the $\phi$-dependence of $\mfD$ because this is inherited from the $\ln\cP$-dependence of $A$, $B$ and $C$, which have been argued to involve subdominant powers of $\alpha_g$. Differentiating with respect to the canonically normalized field $\varphi\simeq \phi/\sqrt{3\cP}$ gives (for real $w_\ssX$) the slow-roll parameters  
\be
\varepsilon = \frac{1}{2}\left(\frac{M_p V_\varphi}{V}\right)^2 = 8 \left(  \frac{M_p w_{\ssX \varphi}}{w_\ssX} \right)^2 \simeq \frac{8\cP}{3} \left(  \frac{4M_p \mfg}{\mft |\phi|^5} \right)^2   \simeq 8 \left( \frac{3}{\cP} \right)^4 \left(  \frac{4M_p \mfg}{\mft |\varphi|^5} \right)^2  \,,
\ee
and
\bea
\eta = \frac{M_p^2V_{\varphi\varphi}}{V} &\simeq&  \frac{4M_p^2 w_{\ssX\varphi\varphi}}{w_\ssX} + 12 \left( \frac{M_p w_{\ssX\varphi}}{w_\ssX} \right)^2   \\
&\simeq& - 4 \left( \frac{3}{\cP} \right)^2 \left(  \frac{20M_p^2 \mfg}{\mft |\varphi|^6} \right) + 12  \left( \frac{3}{\cP} \right)^4 \left(  \frac{4M_p \mfg}{\mft |\varphi|^5} \right)^2   \nn
\eea
where the warping suppression enters through the factor $\mfg/\mft \propto e^{-4\rho}$. 

To see the need for warping it is useful to estimate the size of the factors that enter into $\eta$ and $\varepsilon$. To this end consider the following factor 
\be \label{etaest}
   \eta \ni 4 \left( \frac{3}{\cP} \right)^2 \left(  \frac{20M_p^2 \mfg}{\mft |\varphi|^6} \right) \simeq \frac{80\cP}{3}  \left(  \frac{\mfg}{\mft |\phi|^4} \right)  \left(  \frac{M_p^2}{|\phi|^2} \right) \,.
\ee
Estimating $\mft \sim e^{-2\rho} M_s^2$ and $\mfg \sim e^{-6\rho}M_s^6$ and taking the inter-brane separation to be no bigger than the extra dimensions, which means
\be
   \phi \sim M_s^2 y \lsim M^2_s/M_\KK \sim M_s \cV^{1/6} \sim M_s\, \cP^{1/4} \sim M_p\, \cP^{-1/2}
\ee
then allows the lower bound on \pref{etaest} to be written
\be
   \eta \gsim \frac{80\cP}{3}  \left(  \frac{e^{-4\rho}}{\cP} \right)  \cP  \,.
\ee
Although the requirement $\cP \gg 1$ precludes $\eta$ being small for unwarped geometries \cite{Burgess:2001fx}, a slow roll is possible provided $e^{-4\rho} \ll \cO(\cP^{-1}) \sim \cO(\cV^{-2/3})$, which is consisent with the lower bound $e^{-\rho} \gsim \cV^{-2/3}$ given in \pref{warpbound}.

We see that large enough warping now leads to the predictions $\eta < 0$ and $\varepsilon \ll |\eta| \ll 1$, and this is consistent with the stabilization of the modulus $\tau$ without the $\eta$ problem because it is actually the full quantity $\cP$ that is stabilized. Because $\varepsilon$ is hierarchically smaller than $\eta$ it is clear that the tensor to scalar ratio is so small that there should be no observable primordial tensor fluctuations. If these should be observed in the next few years this inflationary scenario would be decisively ruled out.

As inflation proceeds the value of $w_\ssX$ decreases until at some point the above slow-roll analysis breaks down. Then the fields roll more quickly until they are captured by the minimum or the 4D EFT breaks down and we become unable to predict what happens (such as by having brane and antibrane annihilate and release prodigious amounts of energy). One scenario would be to have $\phi$ reach the zero of $w_\ssX$ before the EFT breaks down (such as in Fig.~\ref{Fig:Coulomb}), in which case the fields can be trapped by the late-time modulus-stabilization solution near $w_\ssX\sim 0$ described in \S\ref{ssec:taustab2}. It is not in general guaranteed that the volume modulus must get trapped at its new minimum, since it can also overshoot and crest the nearby local maximum, followed by a decompactification runaway to infinity. 

Although trapping by the local minimum is not guaranteed, given efficient enough reheating the ruthless efficiency of Hubble friction often makes it much more robust than might naively be expected (as can be seen by numerical evolution in the presence of a thermal background in similar examples \cite{AndyCostasnMe, Conlon:2008cj, YogaDE}). When such trapping occurs the value of the $\tau$ modulus stabilized in its late time minimum $\tau=\tau_0$ can be exponentially larger than its values during inflation. Since $\tau$ determines the sizes of the string and Kaluza-Klein scales relative to the Planck scale such a change can allow the possibility of inflation being controlled by a much larger energy scale than is associated with low-energy supersymmetry breaking and the later universe.

Any large excursion by $\tau$ between inflation and now can easily require the canonical field $\chi \sim M_p \ln\tau$ to run a distance larger than $M_p$. This need not be in contradiction with the distance conjecture \cite{Obied:2018sgi, Palti:2019pca} since the Kaluza-Klein levels provide explicit realizations of the  hypothesized infinite tower of states that descend into (and so ruin) the low-energy theory. The only control issue concerns whether the evolution can be described purely within the low-energy 4D EFT used here. As we have checked, this is easy to ensure during inflation because the fields roll so slowly. It is also fairly easy to arrange for cosmological evolution at later times since the late-time 4D EFT breaks down at scales of order $M_p/\tau_0$ \cite{YogaDE}. Whether a 4D description suffices in between depends somewhat on the nature of the post-inflationary evolution that intervenes between inflation and now, and for some choices of this its evolution might require a more comprehensive UV treatment (such as perhaps along the lines of \cite{Burgess:2016ygs}). 

\subsection{Annihilation and the tachyon superpotential}

We close with more speculative remarks about the small-$\phi$ limit.  Typically the anti D3 brane also hosts other matter fields, including Higgs-like scalar fields $\cH$. These again appear in the low-energy supergravity in a nonsupersymmetric way, appearing in the superpotential only coupled to the goldstino superfield $X$ such as through terms like $W(\cH)=X|\cH|^2$. 

Including this kind of coupling in the superpotential together with the Coulomb interaction gives rise to a superpotential like:\footnote{Recall that $X\ol \cH$ is chiral for a constrained superfield representing a spinless state, allowing terms like $X|\cH|^2$ to appear in the superpotential.}
\be \label{WwTachyon}
W=w_0+Xw_\ssX \quad \hbox{with}\quad w_\ssX= \mft-\frac{\mfg}{|\Phi|^4}-\lambda|\cH|^2 \,.
\ee
In this case the scalar potential is\footnote{Notice that when two fields appear in a potential $V \propto |w_\ssX(\Phi,\cH)|^2$ there is generically a flat direction $\cH(\Phi)$ defined by the condition $w_\ssX(\Phi,\cH)=0$. This direction is generically lifted by the $D$-term potential when the fields carry charge (as would the brane-antibrane tachyon).} (assuming all constants real and positive)
\be
V \propto \frac{|w_\ssX|^2}{\cP^2}=\frac{\left(\mft|\phi|^4-\mfg\right)^2}{\cP^2|\phi|^8}+\frac{2\lambda\left(\mft|\phi|^4-\mfg\right)}{\cP^2|\phi|^4}|\cH|^2+\frac{\lambda^2|\cH|^4}{\cP^2}
\ee
and so as long as the combination $\mft - {\mfg}/{|\phi|^4}$ is positive, the field $\cH$ has a positive mass and the potential is minimized at $\cH = 0$. But once $\phi$ is small enough that $\mft-{\mfg}/{|\phi|^4} < 0$ the canonically normalised field $H \propto \cH/\sqrt{\cP}$ acquires a tachyonic mass 
\be\label{tachymass}
   m^2=\frac{2\lambda}{\cP}\left(  \mft - \frac{\mfg}{|\phi|^4} \right) < 0 \,.
\ee 

\begin{figure}[t]
\begin{center}
\includegraphics[width=100mm,height=60mm]{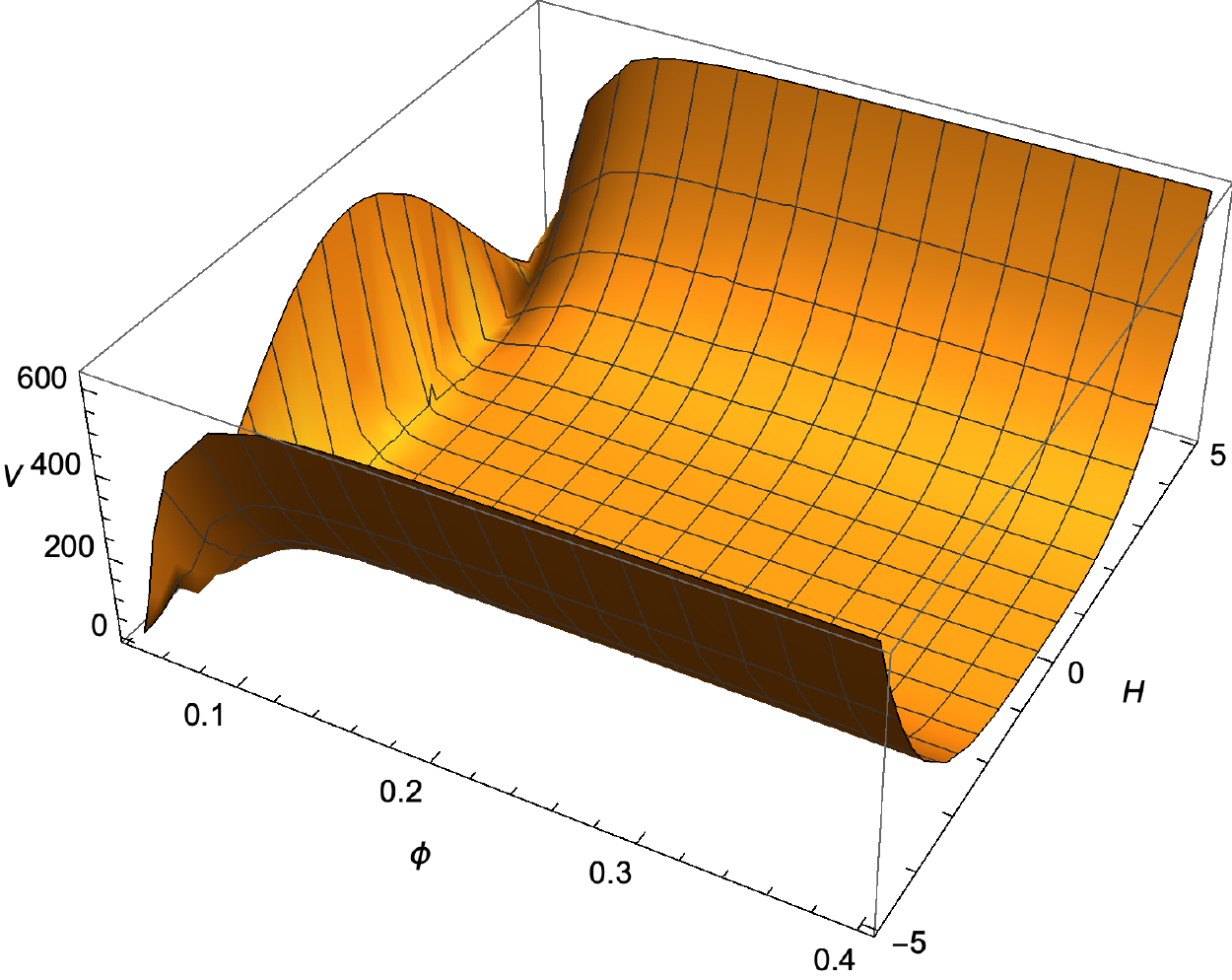} 
\caption{The 4D scalar potential as a function of brane separation $\phi$ and the tachyon $\cH$.} \label{Fig:Tachyon} 
\end{center}
\end{figure}

This captures the same behaviour as would be expected within string theory for a mode of an open string stretching between the brane and the antibrane \footnote{The standard discussion assumes a mass term for the tachyon proportional to $(\phi^2-m^2)\cH^2$ which becomes tachyonic once $\phi$ reaches the mass scale $m$. A similar expression  can be obtained from \pref{tachymass} by expanding $\phi=\phi_0 -\delta\phi $ with $\delta\phi\ll \phi_0$ as done in \cite{Aparicio:2015psl}.}. Such a state becomes lighter as the branes approach one other until at a critical distance it becomes tachyonic (believed to herald the onset of the brane-antibrane annihilation instability). Indeed for large separations, $\phi$, the mass for $\cH$ predicted by \pref{tachymass} becomes proportional to $\mft/\cP \propto e^{-2\rho}/\cP$, which has the same warping and volume dependence as does the square of the warped string scale. See Figure \ref{Fig:Tachyon} for a plot of the potential as a function of $\phi$ and $\cH$, showing in particular a flat inflaton direction with a waterfall-style end of inflation occuring when the $\cH$ field becomes tachyonic.\footnote{We here consider only the simplest potential for the tachyon. Further options can be chosen in order to match the different proposals in the literature \cite{Sen:2004nf}.} Notice that such a tachyonic field is easily `integrated in' within our inflationary picture simply by using \pref{WwTachyon} when evaluating $|w_\ssX|^4$ in the scalar potential \pref{Veffnotau2}. Most of our discussion goes through unchanged because most of our conclusions are independent of the detailed functional dependence of $w_\ssX(\phi)$.  

This simple supersymmetric 4D EFT provides a transparent toy model that captures many features of the full string brane-antibrane annihilation picture \cite{Sen:2004nf}. Since the tachyon field's expectation value breaks the gauge symmetries to which it couples this evolution of $\phi$ to an $\cH$ waterfall provides a dynamical description of symmetry breaking in the antibrane gauge sector.\footnote{The current intuition about the end-point of brane-antibrane annihilation within the full string theory is that no perturbative states remain after tachyon condensation. A puzzle arises because the tachyon field only breaks one combination of the two $U(1)$ gauge symmetries that live on the two branes. It has been conjectured that the second $U(1)$ survives, but in a confining phase that is not manifested perturbatively \cite{Srednicki:1998mq, Yi:1999hd, Sen:1999mg, Gutierrez:2008ya}.  It is tantalising to propose that this late-time behaviour is instead governed by a brane-antibrane bound state -- similar to branonium \cite{Burgess:2003qv} but corresponding to a nontrivial zero of $w_\ssX$ -- rather than continuing to the singularity at $\phi=0$.} Depending on the scales chosen one might build models for which $\cH$ breaks a Grand Unified symmetry at very high scale, or perhaps break the symmetry group of the standard model itself. As noted in \cite{Burgess:2001fx} the tachyon plays other key roles in the physics of the brane-antibrane inflation (besides providing its waterfall finish) such as by giving rise to topological defects such as cosmic strings \cite{Jones:2003da,Copeland:2003bj}. For a discussion on reheating after brane inflation see \cite{Kofman:2005yz}.

\section{Conclusions}

In this paper we explore the consequences of the RG modulus-stabilization mechanism of \cite{AndyCostasnMe, YogaDE} to string vacua, showing that it can generate both de Sitter and non-supersymmetric anti-de Sitter solutions. For the de Sitter vacua no particular uplifting mechanisms is required (unlike for KKLT or LVS stabilization\footnote{It is worth emphasising that even though we introduced anti D3 branes to address inflation, we do not need them to get de Sitter. Furthermore, unlike KKLT, the contribution of the brane tension is only the constant contribution of a field dependent $w_\ssX$ which essentially vanishes at the overall minimum.}). We further explore its implications for relatively simple inflationary scenarios, for which we find it can evade other commonly encountered problems (such as the $\eta$ problem). 

Because the RG mechanism is at heart a perturbative process, it must confront the generic no-go arguments whose roots lie within the old Dine-Seiberg problem \cite{Dine:1985he} (and have been recently revived in the context of swampland conjectures \cite{Obied:2018sgi, Palti:2019pca}). We argue:
\begin{itemize}
\item The Dine-Seiberg issue relies on the scalar potential's runaway behaviour in the weak-coupling limit, and this runaway itself can be interpreted in terms of approximate symmetries of the EFT that are inherited from symmetries of the underlying 10D supergravity (and string theory) underlying the robustness of the problem.  
\item The behaviour of the potential and of evolution in its presence can be addressed within the runaway region using semiclassical expansions without losing calculational control. We argue that the control issues encountered -- such as domain of validity of EFT methods in time-dependent situations for both perturbative and nonperturbative physics -- arise more generally in, and do not undermine conclusions for, other areas of physics and that there is no evidence that string theory effective actions behave differently.
\item Although the Dine-Seiberg argument is true generically -- there are doubtless many solutions in regions of strong coupling ($g_s\simeq \cO(1)$) and small volume ($\tau \simeq \cO(1)$) -- it can be evaded in specific situations.  The RG approach exploits mild hiearchies of coefficients in perturbative expansions of the action (which allows minima to arise with small expansion parameter, $\alpha \ll 1$) together with RG methods that both identify a dependence on $\ln\tau$ and allow controlled resummations that work to all orders in $\alpha \ln \tau$. As a bonus solutions naturally arise at very large values $\ln\tau\simeq 1/\alpha$.
\end{itemize}

Although the RG scenario provides alternatives to the standard KKLT and LVS in IIB string theory, there are plenty of moduli to go around and specific solutions might exploit several of these mechanisms at the same time. For instance, our mechanism may be used to stabilise the overall volume modulus whereas the other K\"ahler moduli may be fixed by non-perturbative effects as in KKLT or LVS. Furthermore, the $\cO(1/10)$ hierarchy required in the RG approach to obtain exponentially large volumes might itself be obtained if these parameters are functions of other fields, such as the complex structure moduli, fixed in other ways. 

There is also nothing intrinsically IIB about the RG mechanism, which can in principle also be put to work in other types of string vacua, such as heterotic models, for in which modulus stabilization has proven to be more challenging (see for instance \cite{Cicoli:2013rwa}) than for IIB. We leave it as an open question to explore this scenario through more explicit constructions, both in IIB string compactifications and in heterotic and type IIA theories.

Combining RG stabilization with string-inflation models can be done by adding a sector with badly broken supersymmetry (to provide the large positive inflationary energy density) and an inflaton field whose evolution slowly interpolates between the early-time large-potential regime and the more negligible scalar potential at later times. The tools of nonlinearly realized supersymmetry and approximate accidental scale invariance are well-suited to the situation where the inflaton is the brane antibrane separation of D3-$\ol{\hbox{D3}}$ inflation scenarios. Because RG stabilization evades the usual $\eta$ problem of these models found in \cite{KKLMMT}, it revives their initial motivation: warping can itself provide the slow roll needed for successful inflation.  Moreover, because the volume modulus takes on different values during and after inflation it allows us to have inflation take place at high (eg GUT) scales and still end at late times with a low enough gravitino mass with TeV supersymmetry breaking in the visible sector.

Although the purely Coulombic interaction potential moves beyond the domain of validity of 4D EFT methods when separations are at the string scale, attractive cosmologies would emerge if the inflaton were to reach a regime with $w_\ssX = 0$ while still in the 4D regime. This is because this naturally becomes a local minimum for which the natural relaxation described more fully in \cite{YogaDE} acts to suppress the value of $V_{\rm min}$ at the minimum. The resulting cosmologies then provide an explicit classical transition between an early inflationary regime and a later de Sitter universe with a suppressed curvature   with the inflaton as the relaxation field.

Finally, we remark how the 4D EFT  with non-linearly realized supersymmetry to which we are led, actually shares many other features of brane-antibrane annihilation as well, such as the transition between relative motion and tachyon condensation. As such it can be regarded as a useful toy model for thinking through what brane annihilation might look like at the end of inflation. This simple scenario is very attractive: it addresses the two main challenges of string inflation -- the $\eta$ problem and the separation of scales while explicitly addressing moduli stabilization -- in a relatively simple framework. We leave to the future the study of concrete string theory models that more fully include all these ingredients.

\section*{Acknowledgements}
We thank Michele Cicoli, Shanta de Alwis, Francesco Muia, Raffaele Savelli, Andreas Schachner, Ashoke Sen and Roberto Valandro for many helpful conversations.  We particularly thank  Michele Cicoli and Shanta de Alwis for useful comments on a preliminary version of this manuscript. CB's research was partially supported by funds from the Natural Sciences and Engineering Research Council (NSERC) of Canada. Research at the Perimeter Institute is supported in part by the Government of Canada through NSERC and by the Province of Ontario through MRI.  The work of FQ has been partially supported by STFC consolidated grants ST/P000681/1, ST/T000694/1. 

\begin{appendix}

\section{Trapping efficiency and effective single-field evolution}
\label{AppTrap}

In the main text we follow \cite{Baumann:2007ah} in assessing the inflationary slow-roll using an approximate single-field model with the volume modulus trapped at a local minimum $\tau = \tau(\phi)$ as the inflaton rolls. In this Appendix we briefly explore the efficacy of this assumption. In principle the gravitational response should be done using the full multi-field model, and it need not be true that an effective single-field description suffices (although it generally does if the non-inflationary field is sufficiently massive). For more detailed discussions of these issues see \cite{Barreiro:2005ua, Panda:2007ie}.

For a general multi-scalar lagrangian
\be
   \cL = - \sqrt{-g} \Bigl[ \frac12 \, G_{ij}(\phi) \, \partial_\mu \phi^i \, \partial^\mu \phi^j + V(\phi) \Bigr]
\ee
the classical equations governing a homogeneous roll are
\be
   \ddot \phi^i + 3H \dot \phi^i + \Gamma^i_{jk} \, \dot \phi^j \, \dot\phi^k + G^{ij} \partial_j V = 0 \,,
\ee 
where as usual 
\be
   \Gamma^i_{jk} := \frac12 \, G^{il} \Bigl( \partial_j G_{kl} + \partial_k G_{jl} - \partial_l G_{jk}\Bigr) 
 \ee
is the Christoffel symbol for the target space metric $G_{ij}$. The energy density and pressure for such a homogeneous roll is
\be
   \rho = \frac12 \, G_{ij} \, \dot \phi^i \, \dot \phi^j  + V(\phi) \quad \hbox{and} \quad
   p = \frac12 \, G_{ij} \, \dot \phi^i \, \dot \phi^j  - V(\phi)  \,.
\ee

In a slow roll the kinetic energy is negligible $\frac12 \, G_{ij} \, \dot \phi^i \, \dot \phi^j  \ll V$ and the evolution is approximately governed by
\be \label{AppSloRoll}
    H^2 = \frac{\rho}3 \simeq \frac{V}3 \quad \hbox{and} \quad
   3H \dot \phi^i  + G^{ij} \partial_j V \simeq  0    \,.
\ee
These equations allow the requirement of small kinetic energy to be cast as a condition on the scalar potential
\be
    \varepsilon := \frac{G^{ij} \partial_i V \partial_j V}{2V^2} \ll 1 \,,
\ee
which also satisfies $\dot H \simeq - \varepsilon H^2$ when time derivatives are evaluated using \pref{AppSloRoll}.

Having ample inflation requires $\varepsilon$ to be small for a sufficiently long time and so also requires its time-derivative to be small. Since using \pref{AppSloRoll} to evaluate the time derivatives of $\varepsilon$ gives
\be
 \frac{\dot \varepsilon}{H} \simeq - \frac{V_{|ij} v^i v^j}{V} + 4 \varepsilon^2 \,,
\ee
where $V_{|ij} := V_{ij} - \Gamma^k_{ij} V_k$ and $v^i := G^{ij} V_j/V$, we see that a field-redefinition invariant representation of the slow-roll conditions $|\dot H/H^2| \ll 1$ and $|\dot \varepsilon/(H\varepsilon)| \ll 1$ is
\be
 \varepsilon = \frac12 G_{ij} v^i v^j \ll 1 \quad \hbox{and} \quad
 \eta_{ij} v^i v^j \ll \varepsilon \quad \hbox{where} \quad \eta_{ij} := \frac{V_{|ij}}{V} \,.
\ee

\subsection*{Brane example}

In the supersymmetric case the target space is a K\"ahler manifold with complex coordinates $\phi^a$ and $\phi^{\bar a}$, for which $G_{ab} = G_{\bar a \bar c} = 0$ and the only nontrivial metric components are $G_{a \bar c} = \partial_a \partial_{\bar c} K$. In this case the only nonzero Christoffel symbols are the purely holomorphic combinations
\be
    \Gamma^a_{bc} = \frac12 G^{\bar e a} \Bigl( \partial_b G_{c \bar e} + \partial_c G_{b \bar e} - \partial_{\bar e} G_{bc} \Bigr) = G^{\bar e a} \partial_b \partial_c \partial_{\bar e} K 
\ee
and their complex conjugates. The evolution for the homogeneous roll of $\phi^a$ is then
\be\label{complexFE}
   \ddot \phi^a + 3H \dot \phi^a + \Gamma^a_{bc} \, \dot \phi^b \, \dot\phi^c + G^{\bar c a} \partial_{\bar c} V = 0 \,.
\ee

For the case considered in the main text we have two complex fields $\phi^a = \{ T , \phi\}$ with $K = - 3 \ln \cP$ with $\cP = T+\ol T - \ol\phi \, \phi$, where $\tau = T + \ol T$ is the real dilaton. In this case the leading derivatives are $K_\ssT = K_{\ol\ssT} = -{3}/{\cP}$, $K_{\ol\phi} = {3\phi}/{\cP}$ and $K_{\phi} = {3\ol\phi}/{\cP}$, and the K\"ahler metric has components
\be
   K_{\ssT\ol\ssT} = \frac{3}{\cP^2} \,, \quad K_{\ssT\ol\phi} = - \frac{3\phi}{\cP^2} \,, \quad K_{\phi\ol\ssT} = - \frac{3\ol\phi}{\cP^2} \quad
   \hbox{and} \quad K_{\phi\ol\phi} = \frac{3}{\cP} + \frac{3 \phi \ol\phi}{\cP^2} = \frac{3\tau}{\cP^2} \,.
\ee
The components of the inverse metric then are
\be
   K^{\ol\ssT\ssT} = \frac{\cP}{3} \Bigl( \cP + \ol\phi \, \phi \Bigr) = \frac{\tau \cP}{3} \,, \quad K^{\ol\ssT\phi} = \frac{\cP \phi}{3} \,, \quad K^{\ol\phi\ssT} =  \frac{\cP \ol\phi}{3}  \quad
   \hbox{and} \quad K^{\ol\phi\phi} = \frac{\cP}{3}   
   \,.
\ee

To compute the Christoffel symbols we require the derivatives $K_{ab\bar c}$, which are
\be
   K_{\ssT\ssT\ol\ssT} = -\frac{6}{\cP^3}  \,, \quad K_{\phi\ssT\ol\ssT} =  \frac{6\ol\phi}{\cP^3}  \,, \quad K_{\phi\phi\ol\ssT} = - \frac{6\ol\phi^2}{\cP^3} 
\ee
and
\be
     K_{\ssT\ssT\ol\phi} = \frac{6\phi}{\cP^3} \,, \quad K_{\ssT\phi\ol\phi} = - \frac{3}{\cP^2} - \frac{6\ol\phi\phi}{\cP^3} \,, \quad K_{\phi\phi\ol\phi} =  \frac{6\tau \ol\phi}{\cP^3} \,, \quad
   \,,
\ee
and so the holomorphic Christoffel symbols are
\be
  \Gamma^\ssT_{\ssT\ssT} = 2\, \Gamma^\phi_{\ssT\phi} = -\frac{2}{\cP} \,, \quad
  \Gamma^\ssT_{\ssT\phi} =  \frac{\ol\phi}{\cP} \,, \quad 
  \Gamma^\phi_{\phi\phi} = \frac{2\ol\phi}{\cP} \quad \hbox{and} \quad
   \Gamma^\phi_{\ssT\ssT} = \Gamma^\ssT_{\phi\phi} =0 \,.
\ee

The field equations \pref{complexFE} for $T$ and $\phi$ therefore become
\be \label{AppTEq}
   \ddot T + 3H \dot T - \frac{2 \dot T^2}{\cP}  + \frac{2\ol\phi \dot \phi \dot T}{\cP} + \frac{\cP \tau V_{\ol\ssT}}{3} + \frac{\cP \ol\phi  V_{\ol\phi}}{3} = 0 \,,
\ee
and
\be \label{AppPhiEq}
   \ddot \phi + 3H \dot \phi + \frac{2 \ol\phi \dot \phi^2}{\cP} - \frac{2 \dot \phi \dot T}{\cP} + \frac{\cP \phi V_{\ol\ssT}}{3} + \frac{\cP  V_{\ol\phi}}{3} = 0 \,.
\ee
Note the absence of the $\dot\phi^2$ terms in the $T$ evolution equation and the absence of the $\dot T^2$ term in the $\phi$ equation. These kinds of term could be dangerous in that {\it e.g.}~$\dot \phi \neq 0$ could become an obstruction to having $\dot T = 0$ even if the potential term in the $T$ equation were to vanish. Such a term tries to drive $T$ along a target-space geodesic, which need not align with the direction towards which the potential encourages the field to move. 

It is clear from \pref{AppPhiEq} that starting at rest near a zero of $V_\ssT$ allows $\phi$ to evolve with the speed expected for the single-field model with $V_{\rm eff}(\phi) = V(\phi, \tau(\phi))$, but the question is whether \pref{AppTEq} also pushes $T$ to evolve so that it remains at its local minimum as $\phi$ evolves. This should occur if the $V_\ssT$ term is the dominant one (for slow motion) in the $T$ equation since then slow roll naturally seeks to adjust $T$ to find the zero of $V_\ssT$.

We can estimate the size of different terms using the potential \pref{VABC} of the main text evaluated in the vicinity of the local minimum $\tau(\phi)$, where all terms --- $w_\ssX^2/\cP^2$, $w_\ssX w_0/\cP^3$ and $w_0^2/\cP^4$ --- are similar in size. This allows the estimates  
\be
   \tau V_\ssT \sim V \sim H^2 \sim \frac{w_0^2}{\cP^4} \quad \hbox{and} \quad
  V_\phi \sim \frac{w_0 w_{\ssX \phi}}{\cP^3} \,.
\ee
In slow roll we therefore expect
\be
  \dot \phi \sim \frac{\cP V_{\ol\phi}}{H} \sim \frac{\cP (w_0 w_{\ssX \phi}/\cP^3)}{w_0/\cP^2} \sim w_{\ssX \phi}
\ee
and so requiring $|\dot\phi|^2/\cP$ be much smaller than $V \sim H^2$ implies $w_{\ssX\phi} \ll {w_0}/{\cP^{3/2}}$. But this makes the $V_\phi \sim w_0 w_{\ssX\phi}/\cP^3 \lsim w_0^2/\cP^{9/2}$ term in the $T$ equation smaller than the $\tau V_\ssT \sim V \sim w_0^2/\cP^4$ term. Because there are also no $\dot \phi^2$ terms this means that an initially motionless $T$ preferentially evolves towards the zero of $V_\ssT$, as required by a single-field treatment. 

Of course real multi-field evolution can be complicated, perhaps oscillating around the trough at $\tau = \tau(\phi)$, depending on the precise initial conditions. When such evolution is studied in detail for brane-antibrane inflation -- as, for example, in \cite{Panda:2007ie} -- it can be the case that the full multi-field evolution gives more inflationary $e$-foldings than would have been inferred using the approximate single-field estimate. 

\end{appendix}

\end{document}